\documentclass[]{aastex631}
\usepackage{graphicx}	
\usepackage{amsmath,bm,amssymb}	
\usepackage{multirow, threeparttable, booktabs, makecell}
\usepackage{hyperref}
\begin{document}

\title[kSZ detection in Fourier space]{\boldmath Detection of pairwise kinetic Sunyaev–Zel’dovich effect with DESI galaxy groups and Planck in Fourier space}

\correspondingauthor{Yi Zheng}
\email{zhengyi27@mail.sysu.edu.cn}

\author{Shaohong Li}
\affiliation{School of Physics and Astronomy, Sun Yat-sen University, 2 Daxue Road, Tangjia, Zhuhai, 519082, China}
\affiliation{CSST Science Center for the Guangdong–Hong Kong–Macau Greater Bay Area, SYSU, Zhuhai 519082, P. R. China}

\author{Yi Zheng}
\affiliation{School of Physics and Astronomy, Sun Yat-sen University, 2 Daxue Road, Tangjia, Zhuhai, 519082, China}
\affiliation{CSST Science Center for the Guangdong–Hong Kong–Macau Greater Bay Area, SYSU, Zhuhai 519082, P. R. China}

\author{Ziyang Chen}
\affiliation{Department of Astronomy, School of Physics and Astronomy, Shanghai JiaoTong University, Shanghai 200240, China}
\affiliation{Key Laboratory for Particle Astrophysics and Cosmology (MOE)/Shanghai Key Laboratory for Particle Physics and Cosmology, China}

\author{Haojie Xu}
\affiliation{Shanghai Astronomical Observatory, Chinese Academy of Sciences, Nandan Road 80, Shanghai 200240, China}
\affiliation{Department of Astronomy, School of Physics and Astronomy, Shanghai JiaoTong University, Shanghai 200240, China}
\affiliation{Key Laboratory for Particle Astrophysics and Cosmology (MOE)/Shanghai Key Laboratory for Particle Physics and Cosmology, China}

\author{Xiaohu Yang}
\affiliation{Department of Astronomy, School of Physics and Astronomy, Shanghai JiaoTong University, Shanghai 200240, China}
\affiliation{Key Laboratory for Particle Astrophysics and Cosmology (MOE)/Shanghai Key Laboratory for Particle Physics and Cosmology, China}
\affiliation{Tsung-Dao Lee Institute, Shanghai Jiao Tong University, Shanghai 200240, China}

\begin{abstract}
We report a $\sim5.2\sigma$ detection of the kinetic Sunyaev–Zel'dovich (kSZ) effect in Fourier space, by combining the DESI galaxy groups and the Planck data. We use the density-weighted pairwise kSZ power spectrum as the summary statistic, and the detailed procedure of its measurement is presented in this paper. Meanwhile, we analyze the redshift space group density power spectrum to constrain its bias parameters and photo-z uncertainties. These best-fitted parameters are substituted to a nonlinear kSZ model, and we fit the measured kSZ power spectrum with this model to constrain the group optical depth $\bar{\tau}$. Selected by a varying lower mass threshold $M_{\rm th}$, the galaxy group catalogs with different median masses ($\tilde{M}$) are constructed from the DR9 data of the DESI Legacy Imaging Surveys. $\tilde{M}$ spans a wide range of $\sim10^{13}-10^{14}M_\odot/h$ and the heaviest $\tilde{M}\sim10^{14} M_\odot/h$ is larger than those of most other kSZ detections. When the aperture photometric filter radius $\theta_{\rm AP}$ is set to be $4.2$ arcmin, the $\tilde{M}=1.75\times10^{13}M_\odot/h$ group sample at the median redshift $\tilde{z}=0.64$ has the highest kSZ detection ${\rm S/N}=5.2$.  By fitting $\bar{\tau}$s from various samples against their $\tilde{M}$s, we obtain a linear $\log\bar{\tau}-\log \tilde{M}$ relation: $\log\bar{\tau} = \gamma(\log \tilde{M}-14)+\log\beta$, in which $\gamma=0.55\pm0.1$. We also vary the aperture photometric filter radius and measure the $\bar{\tau}$ profiles of group samples, whose constraints on the baryon distribution within and around dark matter halos will be discussed in a companion paper.
\end{abstract}


%\begin{document}
%\maketitle
%\flushbottom

\section{Introduction}
\label{sec:intro}
The kinetic Sunyaev-Zel'dovich (kSZ) effect is a secondary cosmic microwave background (CMB) effect induced by the scattering of the CMB photon off free electrons with a bulk motion~\citep{kSZ1970,kSZ1972,kSZ1980,Phillips1995,Birkinshaw1999}. Its detection enables detailed studies of the baryon distribution in the universe and the cosmological model. On one hand, the kSZ signal is linearly proportional to the free electron number density and is independent of the gas temperature. It can be used to constrain the Epoch-of-Reionization (EoR) history~\citep{Alvarez2016,ChenN2023} or the gas distribution within halos~\citep{Schaan16,sugiyama2018,Jonas2021,Schneider2022} and filaments~\citep{ZhengY2023}, and is well suited to search for missing baryons which are believed to mainly reside in low density and low temperature environments. On the other hand, the kSZ effect captures the peculiar velocity field of our universe and records its structure growth information. It can help constrain different aspects of the cosmological model such as Copernican principle~\citep{Zhang11b}, primordial non-Gaussianity~\citep{Munchmeyer2019,Kumar2022}, dark energy~\citep{Pen2014,Okumura22}, modified gravity~\citep{Bianchini2016,Roncarelli2018,Zheng20,Mitchell2021}, massive neutrinos~\citep{Roncarelli2017} and so on.

Albeit, being rich in cosmological information, the detection of the kSZ effect is difficult in its nature. The amplitude of the kSZ signal is 2 orders of magnitudes lower than the CMB primary fluctuation and several times lower than the thermal Sunyaev-Zel'dovich (tSZ) signal~\citep{ZhangP2004}. Besides, its  spectrum is the same as that of the primary CMB; thus, multifrequency measurements do not help on its detection, in contrast to the tSZ effect. As a result, the current detection signal-to-noise ratio (S/N) of the separated kSZ component from the CMB power spectrum is still low~\citep{Dunkley2013,George2015,Bleem2022}. Fortunately, the galaxy resides in high electron density environment and its celestial position denotes where the kSZ signal is relatively high. By cross-correlating the CMB data with the galaxy redshift survey data, we up-weight the high kSZ signal regions and enhance the measurement S/N~\citep{Hand12}. Adopting this so-called kSZ tomography technique, we have detected the late-time (post Epoch-of-Reionization) kSZ effect in a significance around $4\sigma\sim7\sigma$, with combinations of different CMB data and galaxy catalogs, e.g., South Pole Telescope (SPT)+Dark Energy Survey (DES)~\citep{Soergel16,Schiappucci2023}, Atacama Cosmology Telescope (ACT)+SDSS~\citep{Schaan2021,Calafut2021}, Planck/Wilkinson Microwave Anisotropy Probe+unWISE~\citep{Kusiak2021}, Planck+DESI(photo-z)~\citep{Chen2022}, et al. With the next generation of surveys, %such as DESI/Euclid/CSST/PFS/Roman/LSST+Simon/CMB-S4/CMB-HD, 
S/Ns of such measurements can reach $\mathcal{O}(100)$ and beyond~\citep{sugiyama2018,Smith2018}.

The kSZ tomography technique can by implemented by various estimators. Most of them are mathematically equivalent to the three-point correlation function (or bispectrum) of the type $\langle ggT\rangle$ or $\langle gTT\rangle$~\citep{Smith2018}. Here, ``$g$'' is the power of the galaxy field, and ``$T$'' is the power of the CMB field. The majority of such estimators are $\langle ggT\rangle$-like, such as (1) {\it the pairwise estimator}, in which we sum the pair difference of the high-pass filtered CMB at galaxy positions within a certain pair separation bin~\citep{Hand12,sugiyama2018,GongYL2023}; (2) {\it the kSZ template method}, in which the kSZ template is cross-correlated with the CMB map to extract the kSZ signal, and the core of the kSZ template is the projected galaxy momentum field which is reconstructed from the observed galaxy density field~\citep{Ho2009,Shao2011}; (3) {\it the stacked velocity matched filter method}, in which we stack all the high-pass filtered CMB at galaxy locations weighted by the galaxies' velocities to detect the kSZ signal, and the galaxy velocity filed is reconstructed from the galaxy density field~\citep{LiM2014,Schaan16,Schaan2021}; (4) {\it the growth reconstructed method}, in which we directly compare the velocity field constructed from the galaxy density field with the high-pass filtered CMB data at galaxy locations with the maximum likelihood estimation method and extract the structure growth information~\citep{David2016}; (5) {\it the velocity reconstruction method}, in which we directly reconstruct the large-scale radial velocity field from the galaxy density-weighted CMB data with a quadratic estimator~\citep{Deutsch2018} or via the machine-learning technique~\citep{WangYY2021}. The $\langle gTT\rangle$-like estimator includes the one by cross-correlating the large-scale structure with the squared high-pass filtered CMB data~\citep{DeDeo2005,Hill16,Ferraro2016,Patki2023}. Besides these three-point ones, higher order estimators for the kSZ tomography with the $\langle gggT\rangle$-type has also been proposed~\citep{Hernandez2020,Jonas2021,Kuruvilla2022}.

Within all these methods, {\it the pairwise estimator} was used to extract the first cosmological kSZ signal~\citep{Hand12}, and is most widely adopted in a real data analysis~\citep{Hand12,Carlos2015,PlanckkSZ16,Soergel16,DeBernardis17,sugiyama2018,Calafut2021,Chen2022}. We will also implement this estimator in this work.

A dual analysis both in configuration and Fourier space has been a common practice in the galaxy clustering study. Although the correlation function and the power spectrum are Fourier counterparts and contain the same cosmological information in an ideal case, the cosmic variance on large-scales, the shot noise on small scales, and different observational and theoretical systematics restrict the available data within a certain scale range and render different pros and cons to these two statistics~\citep{Verde2007}. For example, the correlation function is straightforward to detect observationally while its covariance matrix estimation and redshift space modeling is nontrivial. On the other hand, the power spectrum has a more diagonal covariance matrix and its anisotropic clustering property is easier to model in redshift space, yet its measurement is contaminated by more systematics such as the anisotropic survey window function and the complicated selection function. In this sense, the analysis in both spaces are complementary to each other and are both valuable in constraining cosmological models. Following this spirit, we implement a Fourier space kSZ analysis in this work, while a complementary configuration space kSZ detection from  the similar data set has been published in~\cite{Chen2022}.

The first and only Fourier space kSZ detection until now is from~\cite{sugiyama2018}.  We follow the main routine developed in that paper and make several modifications/improvements, including the following: (1) the median group masses ($\tilde{M}$) of our samples range from $10^{13}-10^{14}M_\odot/h$, in which the heaviest $\tilde{M}\sim10^{14}M_\odot/h$ is larger than those of most other works, and by this we explore a new range of halo mass in the kSZ detection (see also~\cite{Soergel16}); (2) we implement the aperture photometry (AP) filter in the spherical harmonic space~\citep{Chen2022}, which can avoid the ambiguity of the pixelized temperature average on the circular filter in configuration space; %(3) we try both fixed and non-fixed filter radius, as the clusters in our sample span a large mass and redshift range, a non-fixed filter radius proportional to the cluster angular size guarantees that we are detecting similar regions around clusters and a higher detection S/N; 
(3) {when applying the kSZ power spectrum estimator including the moving line-of-sight (LOS) effect, we decompose the Legendre polynomial into into a product of spherical harmonics~\citep{Hand2017,Sugiyama2018b}, instead of a Cartesian decomposition~\citep{Bianchi2015b,Scoccimarro2015},} and this will reduce the number of needed fast Fouirer transforms (FFTs) if we include the multipole calculation higher than the dipole in the future; {(4) instead of the linear model, we apply a nonlinear kSZ  model developed in~\cite{Zheng20,XiaoL2023} to fit the measured kSZ power spectrum, and since the galaxy group data are from the photometric observation, we further include the photo-z effect in the model; (5) we measure and analyze the redshift space density power spectrum from the same group catalog to obtain constraints on the galaxy bias parameters $b_1$, $b_2$ and photo-z error dispersion $\sigma_{\rm pho-z}$, by which we break the $b_1-\bar{\tau}$ degeneracy in a self-consistent way and avoid possible systematic errors coming from manually selected $b_1-M$ relation and $\sigma_{\rm pho-z}$.}
%fit the measured kSZ power spectrum with a model including the photo-z effect.

The paper is organized as follows. In Section~\ref{sec:data}, we introduce the CMB map and the galaxy group/cluster catalogs we will analyze  in this work. In section~\ref{sec:theory}, a kSZ power spectrum model taking into account the photo-z error is developed. In Section~\ref{sec:method}, we outline the methodology we imply to measure the kSZ power spectrum, and in Section~\ref{sec:result}, we display the main results of this work. We conclude in Section~\ref{sec:conclusion} and more technique details and complementary results are shown in appendices. We will adopt the Planck2018 cosmology~\citep{PLANK2018} throughout the work, and calculations of cosmological quantities are made via the \texttt{Nbodykit} Python toolkit\footnote{\href{https://github.com/bccp/nbodykit}{https://github.com/bccp/nbodykit} from~\cite{Hand2018}.}

\section{Data}
\label{sec:data}
The detection of the pairwise kSZ effect requires the synergy between a CMB map and a galaxy (group/cluster) catalog. The angular position of a galaxy or galaxy group indicates a sky location likely with a high kSZ signal. A compensated AP filter is applied to the CMB data on this location, filtering out the contamination from the primordial CMB fluctuation and the detector noise, and extracting the desired kSZ signal. All these individual kSZ signals are then fed into the pairwise kSZ estimator (Equations~(\ref{eq:cfkSZ}) and~(\ref{eq:PkSZ})) and the output is the statistical pairwise kSZ effect. This pairwise property of the estimator tackles the directional dependence of the kSZ signal since the result simply vanishes in a trivial average of signals from all locations due to the isotropic peculiar velocity field. In this section we first introduce the data sets that we are going to analyze in this work.

\subsection{Planck data}
\label{subsec:planck}
%=============================================fig
\begin{figure}[t!]
\centering
%\subfigure{
\includegraphics[width=0.45\textwidth]{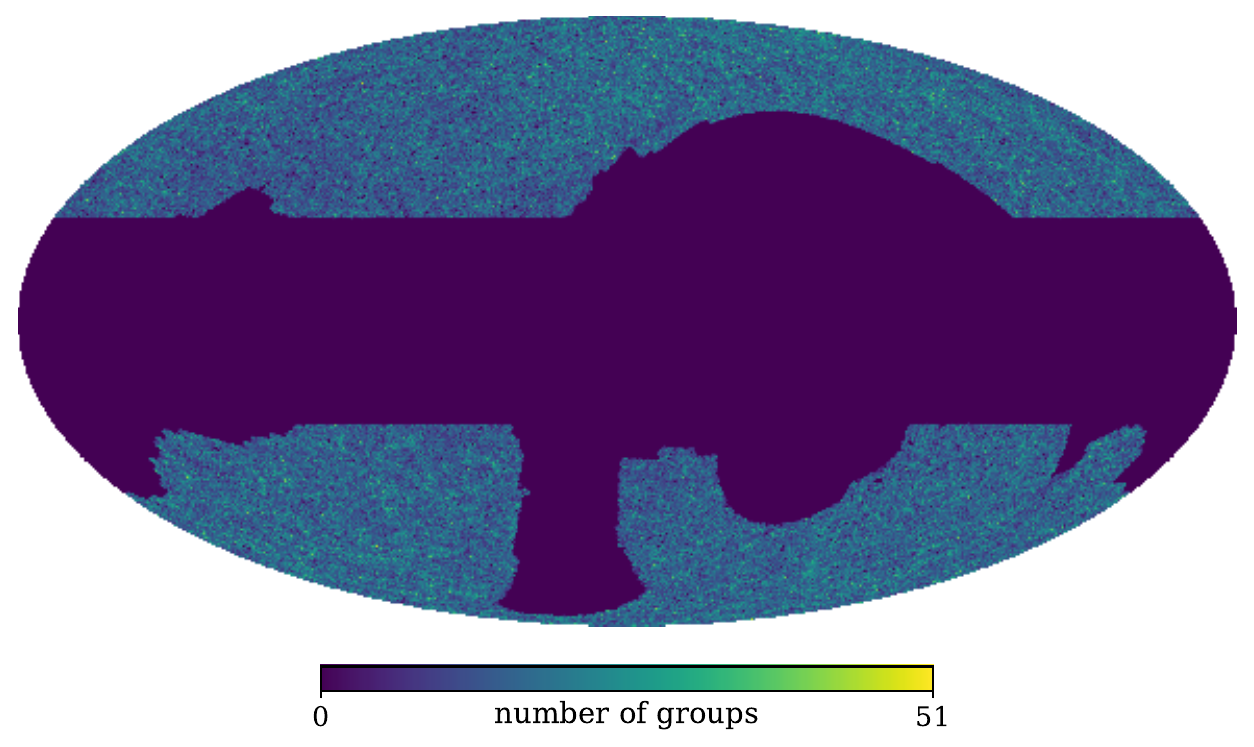}
%}
%\subfigure{
\includegraphics[width=0.45\textwidth]{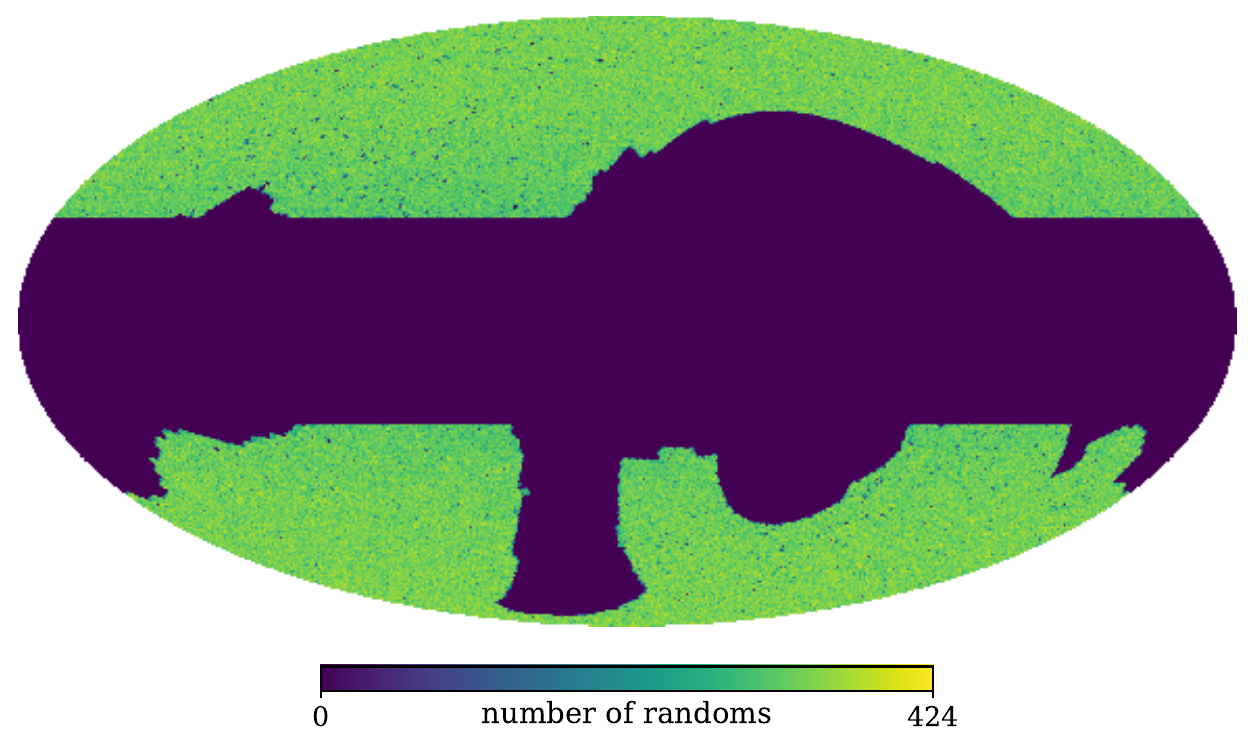}
%}
\caption{\label{fig:sky_distribution}The sky distribution of groups ({\it left}) and random points ({\it right}) of the baseline sample shown in the Galactic coordinate with $N_{\rm side}=256$ in the HEALPix grid frame. %The group catalogue and the corresponding random catalogue are represented with $N_{\rm side}=256$ on the left and right panels respectively. 
The color bar represents the number of groups or random points in each pixel. There are $\sim5.65\times10^6$ groups in the baseline sample, and the random catalog size is 20 times of the corresponding group sample. }
\end{figure}
%=============================================fig
The kSZ signal is measured on a full-sky intensity map, High Frequency Instrument (HFI) 217 GHz$\footnote{\href{https://irsa.ipac.caltech.edu/data/Planck/release_3/all-sky-maps/}{https://irsa.ipac.caltech.edu/data/Planck/release\_3/all-sky-maps/}}$ of the public \textit{Planck} Release 3 data$\footnote{Based on observations obtained with \textit{Planck}(\href{http://www.esa.int/Planck}{http://www.esa.int/Planck}), an ESA science mission with instruments and contributions directly funded by ESA Member States, NASA, and Canada.}$. This map has an effective FWHM $\approx$ 4.87 arcmin and is provided in the HEALPix grid frame~\citep{Gorski2005} with $N_{\rm side}=2048$. We update the map to $N_{\rm side}=8096$ with the zeroth-order interpolation in order to improve the map resolution when applying the AP filter (Section \ref{subsec:AP_filter}). 

The contamination of the tSZ effect is largely alleviated due to its vanishment in the frequency of 217 GHz. Although its residual is nonzero due to the finite bandwidth, it is not expected to have the directional dependence as the pairwise kSZ signal does, thus will be further eliminated in our estimator and does not bias the signal. The same cleaning procedure happens for different foregrounds on the CMB map when we apply the pairwise estimator. Compared to the CMB foreground cleaned maps such as SMICA, SEVEM, NILC and COMMANDER, the 217 GHz map has higher angular resolution, which is highly desired in the kSZ detection. By making comparison between different maps in Appendix C of \cite{Chen2022},  the 217 GHz map is shown to be optimal in our kSZ measurement.

\subsection{DESI photometric Data Release 9 galaxy groups}
\label{subsec:DESIcluster}

%=============================================fig
\begin{figure}[t!]
\centering
\includegraphics[width=0.8\textwidth]{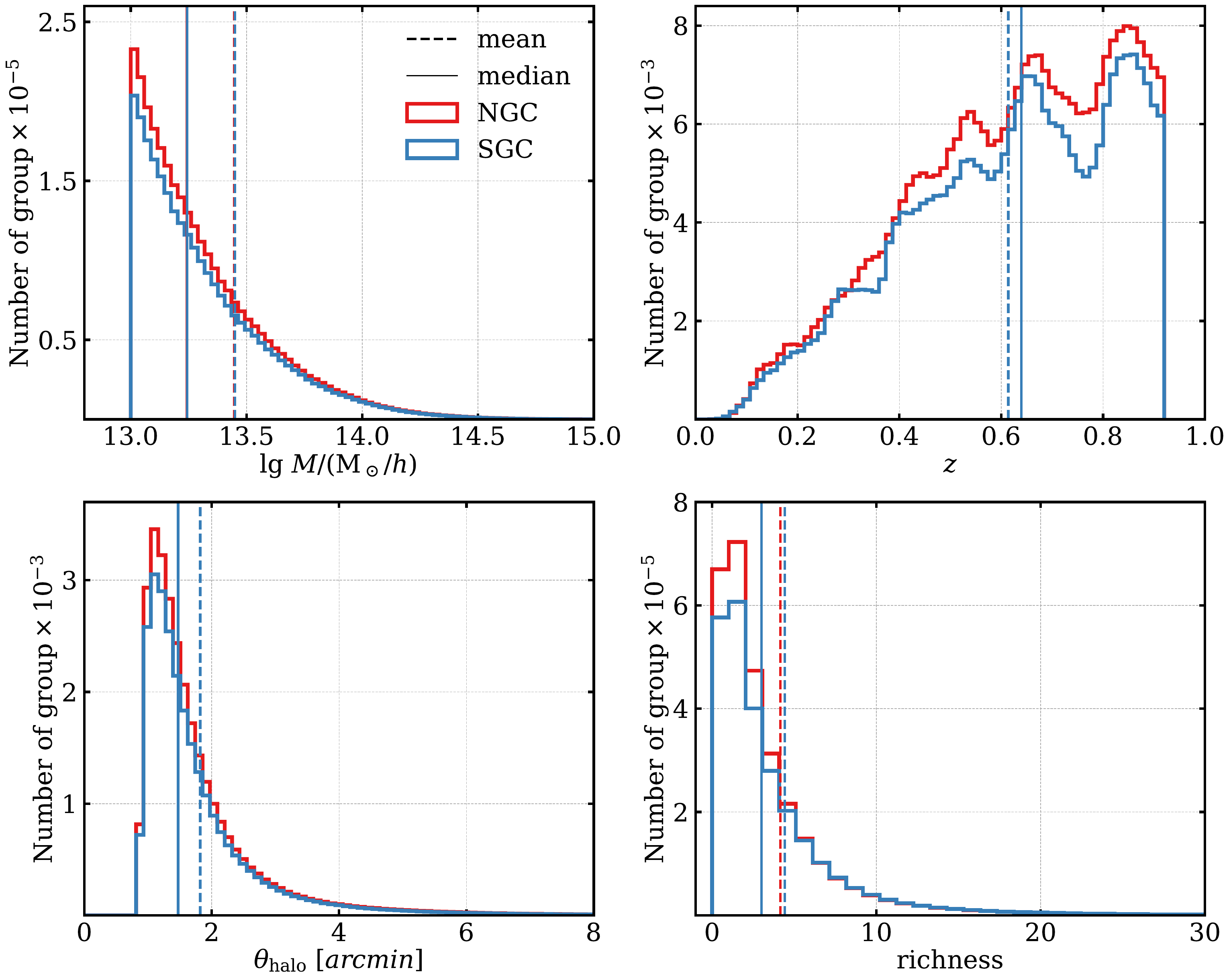}
\caption{\label{fig:mass_z_theta_rich} {The mass, redshift, angular radius and richness distribution of the baseline sample groups. The median values of $\{M,z,\theta_{\rm halo},\rm richness\}$ are $\{1.75\times10^{13}{\rm M_\odot}/h,0.64,1.5',3\}$ for both NGC (red solid lines) and SGC (blue solid lines) groups. The means of these properties are presented by the dashed lines.}}
\end{figure}
%=============================================fig

We use the galaxy group catalog$\footnote{\href{https://gax.sjtu.edu.cn/data/DESI.html}{https://gax.sjtu.edu.cn/data/DESI.html}}$ provided by~\cite{YangXH2021} to identify locations where kSZ signals are large. This catalog is constructed from the Data Release 9 (DR9) of the the DESI Legacy Imaging Survey~\citep{Dey2019}, by extending the halo-based group finder developed in~\cite{YangXH2005}. After removing the sky area within $|b|< 25.^\circ0$ and around nearby sources or masked pixels, the sky coverage of the catalog remains 9622 $\rm deg^2$ in the North galactic cap (NGC) and 8601 $\rm deg^2$ in the South galactic cap (SGC). In total, the group catalog contains 51.44 million galaxy groups in NGC and 45.41 million galaxy groups in SGC with 3D coordinate, richness, halo mass and the total group luminosity. 

The halo gas mass is expected to increase with its total mass (e.g.,~\cite{Chen2022}); therefore, we construct group catalogs by varying a lower group mass threshold $M_{\rm th}$. From this, we construct group samples with median masses ($\tilde{M}$) ranging from $10^{13}-10^{14}M_\odot/h$. In Appendix~\ref{app:diff_sample}, we show how the detection S/N varies with $M_{\rm th}$, and we recognize the group sample with the highest detection S/N  as the baseline sample in this work.

{The baseline sample we construct contains $\sim 5.65\times10^6$ heaviest groups in the group catalog, spanning between $0<z<0.92$\footnote{We truncate the redshift distribution at $z=0.92$ to avoid a spurious redshift distribution peak at $z=0.93$ due to some bugs in the DESI spectroscopic redshifts of the current version.}. Their pixelized sky distribution is shown on the left panel of figure~\ref{fig:sky_distribution}, and their mass, redshift, angular radius and richness distributions are shown in figure~\ref{fig:mass_z_theta_rich}. We mark the median values of $\{M,z,\theta_{\rm halo},\rm richness\}$ for the NGC and SGC subbaseline samples in figure~\ref{fig:mass_z_theta_rich}. The $\{M,z\}$ medians will be used in our following calculation of the theoretical kSZ power spectrum. In Section~\ref{sec:result}, we will present the measured kSZ power spectum and associated results from the baseline sample. 

On the lower left panel of figure~\ref{fig:mass_z_theta_rich}, half of the groups in the baseline sample have an angular radius lower than $1.5'$, much smaller than the Planck FWHM, which results in a heavy smoothing effect on the kSZ signal and limits its S/N. In this sense, the Planck data are not ideal for the SZ effect detection, although its all-sky coverage compensates this disadvantage. On the contrary, high resolution CMB experiments such as ACT and SPT are limited by their relatively small sky coverage overlaps with current galaxy surveys~\citep{Calafut2021,Bleem2022}. On the lower right panel of figure~\ref{fig:mass_z_theta_rich}, half of the groups contain $\le 3$ group members, and most of them reside at high redshifts. The typical photo-z error of the DESI photometric galaxies can be parameterized by $\sigma_{\rm pho-z} = 0.01+0.015z$~\citep{ZhouRP2021,WangK2020,YangXH2021}. For groups with richness $\ge 3$, \cite{YangXH2021} identified that their $\sigma_{\rm pho-z}$ ranges from 0.005 at $z=0$ to 0.015 at $z=1$. These priors indicate that our group samples will have typical photo-z errors of $\sigma_{\rm pho-z}= 0.005\sim0.025$. Such large redshift uncertainty will heavily dilute group clustering along the LOS. Moreover, regarding that our selected samples span large redshift ranges, the complicated photo-z probability distribution function (PDF), which is redshift dependent and possibly asymmetric, makes the modeling of the galaxy group kSZ and density power spectra in redshift space a challenging task. In Appendix~\ref{app:RSD_analysis} we test the sensitivity of the optical depth measurement on different photo-z damping models and make sure our results are not significantly biased by the photo-z uncertainty.}

\section{Theory}
\label{sec:theory}
In this section we present the theoretical model that we will use to fit the measured density-weighted pairwise kSZ power spectra in Section~\ref{sec:result}, which are generated by the methodology described in Section~\ref{sec:method}.

\subsection{kSZ basics}
\label{subsec:basics}
The secondary anisotropy of the CMB temperature caused by the scattering off of CMB photons by free electrons with bulk motions is known as the kSZ effect~\citep{kSZ1972,kSZ1980}. The change of the CMB temperature is 
\begin{equation}
\label{eq:delta_T}
    \begin{split}
    \frac{\Delta T_{\rm kSZ}}{ T_{\rm CMB}}(\hat{\boldsymbol{n}}) =-\int{n_e(\hat{\boldsymbol{n}},l)\sigma_{\rm T}\frac{\boldsymbol{v}\cdot\hat{\boldsymbol{n}}}{c}}dl\,.
    \end{split}
\end{equation}
Here, $n_e$ is the physical free electron number density, $\boldsymbol{v}\cdot\hat{\boldsymbol{n}}$ is the electron peculiar velocity projected along the LOS direction, $T_{\rm CMB}=2.7255K$ is the average CMB temperature, $\sigma_{\rm T}$ is the Thomson scattering cross section, and $c$ is the speed of light. The integral $\int dl$ in this expression is performed along the LOS. 

If the CMB photons coming from the $\hat{\boldsymbol{n}}$ direction are only scattered once by the $i$th galaxy group, the kSZ effect generated by the cloud of free electrons around this group can be simplified as
\begin{equation}
\label{eq:delta_T_i}
    \begin{split}
    \frac{\Delta T_{i,\rm kSZ}}{ T_{\rm CMB}}(\hat{\boldsymbol{n}}_i) =-\tau_i\frac{\boldsymbol{v}_i\cdot\hat{\boldsymbol{n}}_i}{c}\,,
    \end{split}
\end{equation}
where $\boldsymbol{{v}_i}$ is the physical bulk motion of the $i$th group, and $\tau_i=\int n_e\sigma_{\rm T}dl$ is its optical depth, depending on its baryon abundance. Assuming further that all groups share approximately the same optical depth, we have
\begin{equation}
\label{eq:delta_T_i2}
    \begin{split}
    \frac{\Delta T_{i,\rm kSZ}}{ T_{\rm CMB}}(\hat{\boldsymbol{n}}_i) =-\bar{\tau}\frac{\boldsymbol{v}_i\cdot\hat{\boldsymbol{n}}_i}{c}\,,
    \end{split}
\end{equation}
in which $\bar{\tau}$ is the average optical depth of the group sample.

\subsection{Density-weighted pairwise kSZ power spectrum}
\label{subsec:Pairwise_kSZ}

The trivial average of the kSZ signal in Equation~(\ref{eq:delta_T_i2}) vanishes due to its directionl dependence.
%Due to the direction dependence shown in Equation~(\ref{eq:delta_T_i2}), the kSZ signal vanishes in a trivial average of $\Delta T_{i,\rm kSZ}$'s. 
Fortunately, under the influence of gravity, there is a tendency of two groups moving toward each other on quasi to nonlinear scales. In turn, we can detect the kSZ effect by a pairwise estimator~\citep{sugiyama2018}:
\begin{equation}
\label{eq:cfkSZ}
    \begin{split}
        \xi_{\rm kSZ}(\boldsymbol{s})=\left\langle-\frac{V}{N^2}\sum_{i,j}\left[ \Delta T_{i,\rm kSZ}-\Delta T_{j,\rm kSZ}\right]\delta_D(\boldsymbol{s}-\boldsymbol{s}_{ij}) \right\rangle\,,
    \end{split}
\end{equation}
where $\xi_{\rm kSZ}$ is the density-weighted pairwise kSZ correlation function, $\boldsymbol{s}_{ij}=\boldsymbol{s}_{i}-\boldsymbol{s}_{j}$ where $\boldsymbol{s}_{i}$ denotes the redshift space position of the $i$th group, $N$ represents the number of observed groups, and $V$ is the volume of the survey. 

The Fourier counterpart of $\xi_{\rm kSZ}$ is the density-weighted pairwise kSZ power spectrum,
\begin{equation}
\label{eq:PkSZ}
    \begin{split}
    P_{\rm kSZ}(\boldsymbol{k})=\left\langle-\frac{V}{N^2}\sum_{i,j}\left[ \Delta T_{i,\rm kSZ}-\Delta T_{j,\rm kSZ}\right]e^{-i\boldsymbol{k}\cdot{}\boldsymbol{s}_{ij}} \right\rangle\,.
    \end{split}
\end{equation}
Inserting Equation~(\ref{eq:delta_T_i2}) into Equations~(\ref{eq:cfkSZ}) and~(\ref{eq:PkSZ}), $\xi_{\rm kSZ}$ and $P_{\rm kSZ}$ are proportional to the density-weighted pairwise velocity correlation function $\xi_{\rm pv}$ and power spectrum $P_{\rm pv}$ respectively: 
\begin{equation}
\label{eq:xikSZ_to_pv}
    \begin{split}
    \xi_{\rm kSZ}(\boldsymbol{s})\simeq \frac{T_{\rm CMB}\bar{\tau}}{c}\xi_{\rm pv}(\boldsymbol{s})\,,
    \end{split}
\end{equation}
\begin{equation}
\label{eq:PkSZ_to_pv}
    \begin{split}
    P_{\rm kSZ}(\boldsymbol{k})\simeq \frac{T_{\rm CMB}\bar{\tau}}{c}P_{\rm pv}(\boldsymbol{k})\,,
    \end{split}
\end{equation}
where
\begin{equation}
\label{eq:Ppv}
    \begin{split}
    \xi_{\rm pv}(\boldsymbol{s})=\left\langle-\frac{V}{N^2}\sum_{i,j}\left[ \boldsymbol{v}_i\cdot{}\hat{\boldsymbol{n}}_i- \boldsymbol{v}_j\cdot{}\hat{\boldsymbol{n}}_j\right]\delta_D(\boldsymbol{s}-\boldsymbol{s}_{ij}) \right\rangle\,,
    \end{split}
\end{equation}
\begin{equation}
\label{eq:xipv}
\begin{split}
        P_{\rm pv}(\boldsymbol{k})=\left\langle-\frac{V}{N^2}\sum_{i,j}\left[ \boldsymbol{v}_i\cdot{}\hat{\boldsymbol{n}}_i- \boldsymbol{v}_j\cdot{}\hat{\boldsymbol{n}}_j\right]e^{-i\boldsymbol{k}\cdot{}\boldsymbol{s}_{ij}} \right\rangle\,.
    \end{split}
\end{equation}
A hidden assumption adopted here is that the group optical depth has no correlation with the group peculiar velocity. 

Compared to the traditional pair-weighted statistics, e.g., $\xi_{\rm pv}^{\rm pair-weight} = \xi_{\rm pv}/(1+\xi_{\delta})$, the density-weighted statistics encode similar cosmological information, while their modeling is simpler by avoiding the modeling of an additional density correlation function $\xi_\delta$ in the denominator. A detailed comparison between the density- and pair-weighted statistics can be found in Appendix B of \cite{Okumura2014}.

Under the plane-parallel approximation and the condition that the peculiar velocity field is proportional to the linear growth rate $f\equiv d\ln D/d\ln a$ (which is generally true and $D$ is the linear growth factor), $P_{\rm pv}$ in redshift space can be derived from the redshift space galaxy power spectrum $P_{\rm s}(\boldsymbol{k})$ by \citep{Sugiyama2016,Sugiyama2017} 
\begin{equation}
\label{eq:Ppv_to_Ps}
    \begin{split}
    P_{\rm pv}(\boldsymbol{k})=\left( i\frac{aHf}{\boldsymbol{k}\cdot{}\hat{\boldsymbol{n}}}\right)\frac{\partial}{\partial f}P_{\rm s}(\boldsymbol{k})\,,
    \end{split}
\end{equation}
where $a$ is the cosmic scale factor, $H$ is the Hubble parameter.

{In this work, we use the following nonlinear galaxy power spectrum to derive the kSZ power spectrum: %derived from the linear $P_{\rm s}(\boldsymbol{k})$ to fit the measurement, namely~\citep{Kaiser87}
\begin{equation}
\label{eq:nonlinear_Ps}
    \begin{split}
        P^{\rm spec-z}_{\rm s}(k,\mu)=\left(b_1+b_2k^2e^{-k^2}+f\tilde{W}\mu^2\right)^2P_{\rm m}(k)\frac{1}{1+k^2\mu^2\sigma_v^2/H^2}\,.
    \end{split}
\end{equation}
In Appendix~\ref{app:RSD_analysis}, this model is adopted to analyze the redshift space group density power spectrum. Here, the superscript `spec-z' suggests that we are analyzing a galaxy sample with negligible redshift uncertainties, e.g., from a spectroscopic galaxy survey, $b_1$ and $b_2$ are linear and nonlinear galaxy density bias parameters, $\mu\equiv \cos\theta$ denotes the cosine of the angle $\theta$ between $\boldsymbol{k}$ and the LOS, $P_{\rm m}(k)$ is the nonlinear dark matter power spectrum at redshift $z$, a Lorentzian Finger-of-God (FoG) term is adopted and the matter velocity dispersion $\sigma_v^2$ is calculated by  $\sigma^2_v = \int f^2H^2P_{\rm m}dk/6\pi^2$\footnote{In the spec-z data redshift space distortion (RSD) analysis, $\sigma_v^2$ is usually treated as a free parameter, which effectively absorbs the uncertainty of the Kaiser term modeling. In this photo-z data analysis, the power spectrum damping from the photo-z error is much larger than that induced by $\sigma_v^2$. So when we treat the photo-z error dispersion $\sigma^2_{\rm pho-z}$ as a free parameter, a direct calculation of $\sigma_v^2$ here will not trigger severe systematic errors.}, assuming a unity volume-weighted galaxy velocity bias~\citep{Zheng14b,Junde18vb}.

In order to tackle the nonlinear correlation between density and velocity fields, we include the window function $\tilde{W}(k)\equiv P_{\delta\theta}/fP_{\delta\delta}$ in Equation~(\ref{eq:nonlinear_Ps})~\citep{Zhangrsd}. Motivated by the perturbation theory and verified with N-body simulations, $\tilde{W}(k)$ can be described by the following fitting function~\citep{Zheng13}
\begin{equation}
\label{eq:W}
\tilde{W}(k) = \frac{1}{1+\Delta\alpha(z)\Delta^2_{\delta\delta}(k,z)}\,,
\end{equation}
where $\Delta^2_{\delta\delta}(k,z) = k^3P_{\delta\delta}/2\pi^2$ is the dimensionless density power spectrum, and the redshift dependence of $\Delta\alpha$ can be fitted well by a linear relation~\citep{Zheng13}
\begin{equation}
\label{eq:Delta_alpha}
\Delta\alpha(z) = \Delta\alpha(z=0)+0.082z = 0.331+0.082z\,.
\end{equation}

By inserting Equation~(\ref{eq:nonlinear_Ps}) into Equations~(\ref{eq:PkSZ_to_pv}) and~(\ref{eq:Ppv_to_Ps}), we derive the nonlinear density-weighted pairwise kSZ power spectrum as
\begin{eqnarray}
\label{eq:theoretical_PkSZ_nonlinear}
    % \begin{split}
P^{\rm spec-z}_{\rm kSZ}(k,\mu)&=&\left(\frac{T_{\rm CMB}\bar{\tau}}{c}\right)2iaHf\mu(b_1+b_2k^2e^{-k^2}+f\tilde{W}\mu^2)\frac{P_{\rm m}(k)}{k} S_{\rm L}(k,\mu) \frac{1}{1+k^2\mu^2\sigma_v^2/H^2} \,,\\
S_{\rm L}(k,\mu)&=&\tilde{W}-[(b_1+b_2k^2e^{-k^2})/f+\tilde{W}\mu^2]\frac{k^2\sigma_v^2/H^2}{1+k^2\mu^2\sigma_v^2/H^2}\,.\nonumber
    % \end{split}
\end{eqnarray}

At linear scales, the above nonlinear model can be simplified to a linear one
\begin{equation}
\label{eq:theoretical_PkSZ_lin}
    \begin{split}
        P^{\rm spec-z}_{\rm kSZ}(k,\mu)=\left(\frac{T_{\rm CMB}\bar{\tau}}{c}\right)2iaHf\mu(b_1+f\mu^2)\frac{P_{\rm m}(k)}{k}\,.
    \end{split}
\end{equation}
It is easy to figure out the degeneracy between $b_1$ and $\bar{\tau}$, namely that an accurate $\bar{\tau}$ detection requires an accurate $b_1$ constraint. In the next section, we further notice that the photo-z uncertainty heavily damps the kSZ power spectrum, and this damping is also degenerated with $b_1$ and $\bar{\tau}$. This is a unique feature for the photo-z data. In Appendix~\ref{app:RSD_analysis}, we tackle these issues by applying a joint analysis of kSZ and RSD analysis on the same data set. The tests there show that our $\bar{\tau}$ measurements are minimally biased by the abovementioned degeneacies.
}

\subsection{Density-weighted pairwise kSZ power spectrum with photometric redshits}
\label{subsec:Pairwise_kSZ_photo}
%=============================================fig
\begin{figure}[t!]
\centering
\includegraphics[width=0.6\textwidth]{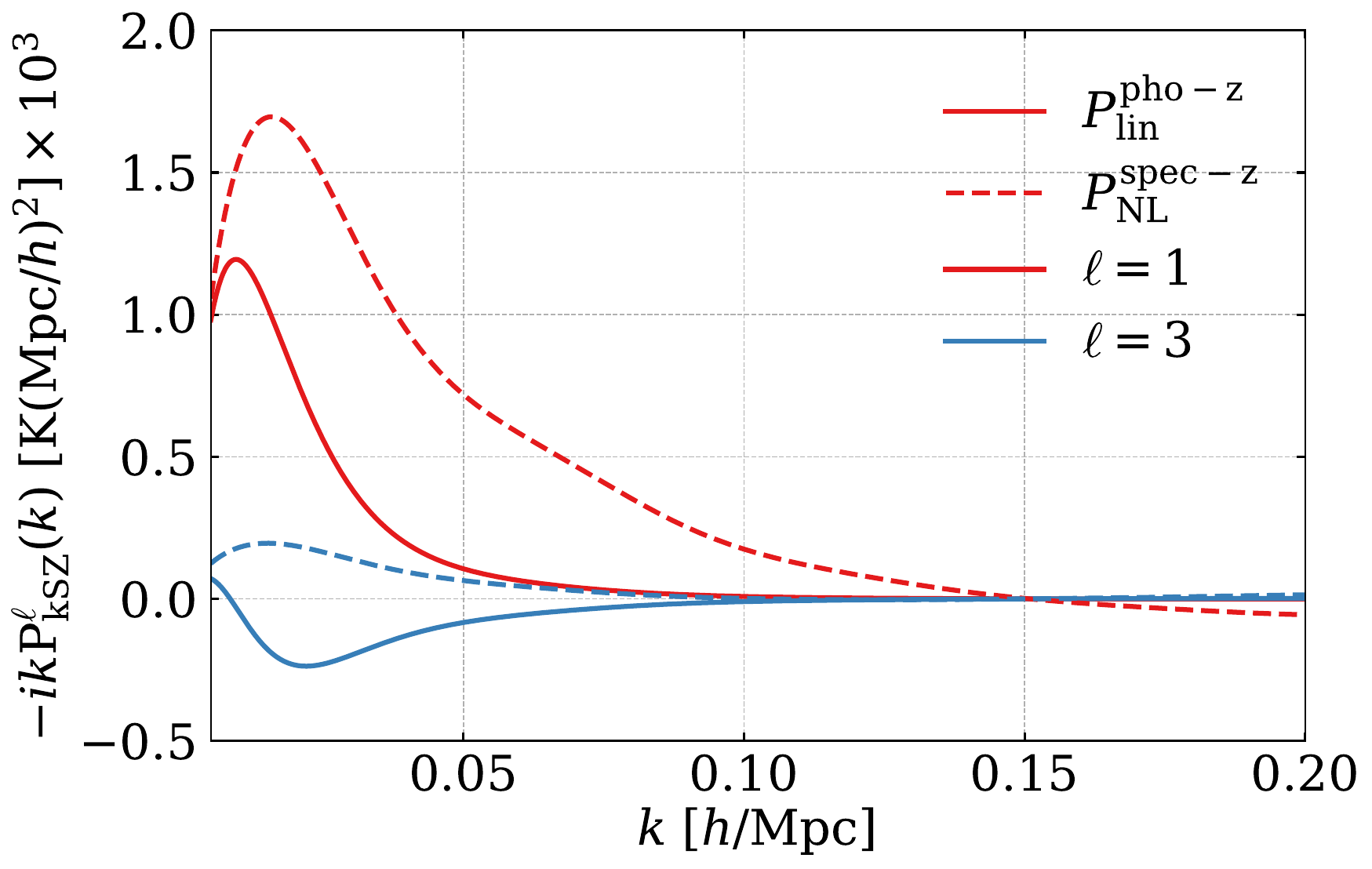}
\caption{\label{fig:Pkl_kSZ_theory} {Theoretical predictions of the unmasked pairwise kSZ multipoles of the NGC sub-sample of the baseline catalog. Solid lines represent multipoles of the nonlinear density-weighted pairwise kSZ power spectrum with photo-z errors (Equation~(\ref{eq:theoretical_PkSZ_photoz})) and dashed lines are those without photo-z errors (Equation~(\ref{eq:theoretical_PkSZ_nonlinear})). The red and blue lines are for dipoles ($\ell=1$) and  octopoles ($\ell=3$) respectively. The adopted redshift is the median redshift $\tilde{z}=0.64$ of the NGC sub-sample, and $\bar{\tau}=3.96\times10^{-5}$ is the best-fitted $\bar{\tau}$ value of the baseline sample.}}
\end{figure}
%=============================================fig

The photometric data we analyze have nonnegligible redshift errors. These photo-z errors randomly distort the LOS positions of galaxy groups and dilute their LOS clustering, which is an analogy of the FoG effect. {As discussed in detail in Appendix~\ref{app:RSD_analysis}, we model the distribution function of the pairwise photo-z errors as an exponential function with zero mean and a dispersion of $\sigma^{\rm pair}_{\rm pho-z}=\sqrt{2}\sigma_{\rm pho-z}$, with $\sigma_{\rm pho-z}$ being the dispersion of the 1-D photo-z error distribution function; then, the galaxy power spectrum of the photo-z data becomes~\citep{Davis1983,Ballinger96}
\begin{eqnarray}
\label{eq:theoretical_Ps_photoz}
    P_{\rm s}^{\rm pho-z} &=& P^{\rm spec-z}_{\rm s}D_{\rm pho-z} \nonumber\\
    &=& P^{\rm spec-z}_{\rm s}\frac{1}{1+k^2\mu^2\sigma_z^2/H^2}\,,
\end{eqnarray}
where $D_{\rm pho-z}$ is the photo-z damping function and $\sigma_z\equiv c\sigma_{\rm pho-z}$.  Since $\sigma_z$ does not depend on $f$, the resultant kSZ power spectrum is similarly diluted by the photo-z errors as
\begin{equation}
\label{eq:theoretical_PkSZ_photoz}
    P_{\rm kSZ}^{\rm pho-z} = P^{\rm spec-z}_{\rm kSZ}\frac{1}{1+k^2\mu^2\sigma_z^2/H^2}\,.
\end{equation}
We will omit the superscript `$\rm pho-z$' hereafter.

In Appendix~\ref{app:RSD_analysis}, we measure the galaxy density power spectrum multipoles of group catalogs and fit the measurements with the nonlinear model in Equations~(\ref{eq:nonlinear_Ps}) and~(\ref{eq:theoretical_Ps_photoz}). With $f$ being fixed to the fiducial cosmological value, this fitting determines the posterior distribution of $b_1$, $b_2$ and $\sigma_z$. We then substitute the best-fitted values of these 3 parameters to Equation~(\ref{eq:theoretical_PkSZ_nonlinear}) and fit the kSZ measurement to obtain the constraint on $\bar{\tau}$. In this way, (1) we model the galaxy density and kSZ power spectra in a self-consistent manner; (2) we self-consistently break the $b_1-\bar{\tau}$ degeneracy and avoid possible systematic biases coming from a manually selected $b_1-M$ relation and $\sigma_{\rm pho-z}$.; (3) as $\bar{\tau}$ is the only parameter in the kSZ measurement fitting, the S/N of the fitted $\bar{\tau}$ directly represents the S/N of the kSZ detection, which is one of main targets of this work. For a global fitting of the galaxy density and kSZ power spectra simultaneously, which requires calculations of the covariance matrix between two power spectra, we defer it to the future work.
}

Moreover, we expand $P_{\rm kSZ}(k,\mu)$ in terms of Legendre polynomials, %(see \citep{sugiyama2018} and references therein,and more detail in \citep{Alam2017}), 
\begin{equation}
\label{eq:theoretical_PkSZ_l}
    \begin{split}
    P_{\rm kSZ}^{\ell}(k)=\frac{2\ell+1}{2}\int^1_{\rm -1}d\mu\mathcal{L}_\ell(\mu)P_{\rm kSZ}(k,\mu)\,.
    \end{split}
\end{equation}
%where $d\Omega_k$ is the solid angle element in $k$-space.
Since $P_{\rm kSZ}(k,\mu)$ is an odd function on $\mu$,  $P^{\ell}_{\rm kSZ}$ is  nonzero only for odd $\ell$s, in contrast to the galaxy power spectrum $P_{\rm s}(k,\mu)$. As shown by solid lines in figure~\ref{fig:Pkl_kSZ_theory},  the absolute values of $P^{\ell=3}_{\rm kSZ}(k,\mu)$ are smaller than those of $P^{\ell=1}_{\rm kSZ}(k,\mu)$, and in principle, the measurement errors of the former will be larger. Considering the limited S/N of our measurement, we only focus on the dipole $P^{\ell=1}_{\rm kSZ}$ in this work. 

{By dashed lines, we also plot the nonlinear kSZ power spectrum multipoles without photo-z errors evaluated by Equation~(\ref{eq:theoretical_PkSZ_nonlinear}) in figure~\ref{fig:Pkl_kSZ_theory}.} These will be signals likely detected by future spectroscopic galaxy surveys, e.g., DESI~\citep{DESI2016} or Subaru Prime Focus Spectrograph (PFS)~\citep{PFS12}. Without the smearing effect of photo-z errors, kSZ multipoles exhibit more measurable features at nonlinear scales, and will guarantee a kSZ detection of higher S/N~\citep{sugiyama2018}.

\section{Methodology}
\label{sec:method}
In this section, we outline the methodology that we follow to measure the kSZ signal and the associated covariance matrix.

\subsection{Aperture photometry filter}
\label{subsec:AP_filter}

In order to minimize the contamination from the primary CMB, an AP filter is applied to the Planck CMB map (e.g.~\cite{sugiyama2018,Calafut2021,Chen2022}). The AP filter is a 2D compensated top-hat filter:

\begin{equation}
\label{eq:W_AP_theta}
    \begin{split}
    W_{\rm AP}(\theta)=\frac{1}{\pi \theta_{\rm AP}^2}\left\{
        \begin{aligned}
            1,\ \ \ \ \ \ \ \ \ \ \ \ \ \ \ \ \  \theta \le \theta_{\rm AP}\,,\\    
            -1,\ \ \ \theta_{\rm AP}< \theta \le \sqrt{2}\theta_{\rm AP}\,,\\
            0,\ \ \ \ \ \ \ \ \ \ \ \ \   \theta > \sqrt{2} \theta_{\rm AP}\,,
        \end{aligned}
    \right.
    \end{split}
\end{equation}
where the averaged CMB temperature within a disk of aperture size $\theta_{\rm AP}$ and an annulus of equal area, out to radius $\sqrt{2} \theta_{\rm AP}$, are differentiated around each galaxy group to measure the kSZ signal $\Delta T^{\rm AP}_{\rm kSZ}$:
\begin{equation}
\label{eq:delta_T_AP}
    \begin{aligned}
        \Delta T_{\rm kSZ}^{\rm AP}(\boldsymbol{\theta})= \int d^2 \theta' W_{\rm AP}(\boldsymbol{\theta}-\boldsymbol{\theta}')T_{\rm map}(\boldsymbol{\theta}')\,,
    \end{aligned}
\end{equation}
in which $T_{\rm map}$ denotes the CMB temperature map.

In this work, we follow~\cite{Chen2022} and apply the AP filter in the spherical harmonic space, namely 
\begin{equation}
\label{eq:delta_T_AP_l}
    \Delta T_{\rm kSZ}^{\rm AP}(\boldsymbol{\theta})=\int\frac{d^2\ell}{(2\pi)^2}e^{i\boldsymbol{\ell}\cdot\boldsymbol{\theta} }W_{\rm AP}(\ell\theta_{\rm AP})T_{\rm map}(\boldsymbol{\ell})\,,
\end{equation}
where
\begin{equation}
\label{eq:W_AP_l}
    \begin{split}
    W_{\rm AP}(\ell\theta_{\rm AP})=\frac{2}{\ell\theta_{\rm AP}}[2J_1(\ell\theta_{\rm AP})-\sqrt{2}J_1(\sqrt{2}\ell\theta_{\rm AP})]
    \end{split}
\end{equation}
and $J_1$ is the first Bessel function of the first kind.

\subsection{Random catalog}
\label{subsec:random}

We rely on the random catalog to calculate the galaxy group density field in Equation~(\ref{eq:delta_n_3D}). We generate a corresponding random catalog for each group sample we construct, by randomly downsizing the random catalog provided along with the DESI Legacy Imaging Survey DR9 data\footnote{\href{https://portal.nersc.gov/cfs/cosmo/data/legacysurvey/dr9/randoms/}{https://portal.nersc.gov/cfs/cosmo/data/legacysurvey/dr9/randoms/}}. The size of each random catalog is 20 times of the size of the corresponding group sample. Since the group catalog of~\cite{YangXH2021} is constructed within the continuous region of the DESI sky coverage, we manually discard the discontinuous regions in the random catalog in the first step. The pixelized sky distribution of the random catalog is shown on the right panel of figure~\ref{fig:sky_distribution}.

We apply the shuffled method \citep{Ross2012}, which is proven to be robust in the galaxy clustering analysis, to assign redshifts to the random catalog. In this method, each random point is assigned a redshift which is randomly selected from the redshifts of the group sample.

%The random catalogue$\footnote{\href{https://portal.nersc.gov/cfs/cosmo/data/legacysurvey/dr9/randoms/}{https://portal.nersc.gov/cfs/cosmo/data/legacysurvey/dr9/randoms/}}$ of DR9 with random distribution in the sky survey area characterize the coverage of galaxies. The conditions and flags, applied in the selections of galaxies by \citep{Yang2021}, are employed to obtain the same area as groups. The HEALPix with $N_{\rm side}=128$ is used to remove random objects obviously not contained in the group area. The number of objects of random catalog is randomly selected to be 100 times as large as group sample. We use suffled method (referring to section 6.1 of \citep{Ross2012}) to produce the redshifts of random catalogue.

\subsection{kSZ and galaxy group density fluctuations}
\label{subsec:density_field}

In Section \ref{subsec:AP_filter}, the AP filtered CMB temperature $\Delta T_{\rm kSZ}^{\rm AP}$ of each group is estimated in the spherical harmonic space. To suppress the systematic errors induced by the redshift-dependent foregrounds such as the cosmic infrared background and the residual tSZ signal, we subtract from $\Delta T_{\rm kSZ}^{\rm AP}$ its average  with a Gaussian weight \citep{Hand12}:
\begin{equation}
\label{eq:T}
    \begin{split}
        \Delta T_{\rm kSZ}(\hat{\boldsymbol{n}}_i)= \Delta T_{\rm kSZ}^{\rm AP}(\hat{\boldsymbol{n}}_i) -\frac{\sum_j  \Delta T_{\rm kSZ}^{\rm AP}(\hat{\boldsymbol{n}}_j)G(z_i,z_j,\sigma_{z})}{\sum_j G(z_i,z_j,\sigma_{z})}\,,
    \end{split}
\end{equation}
where $G(z_i,z_j,\sigma_{z}) ={\rm exp}(-z^2_{ij}/2\sigma^2_{z})$ with $\sigma_z= 0.01$, and $z_{ij}\equiv z_i-z_j$. The 3D kSZ temperature fluctuation field then reads
\begin{equation}
\label{eq:delta_T_3D}
    \begin{aligned}
        \delta T (\boldsymbol{s})&=\sum^{N_{\rm group}}_i \Delta T_{\rm kSZ}(\hat{\boldsymbol{n}}_i)\delta_{\rm D}(\boldsymbol{s}-\boldsymbol{s}_i)\,,
    \end{aligned}
\end{equation}
where $N_{\rm group}$ is the size of the group sample and $\delta_{\rm D}$ is the 3D Dirac delta function.

With the help of the random catalog constructed in Section~\ref{subsec:random}, we define the 3D galaxy group density field as
\begin{equation}
\label{eq:delta_n_3D}
    \begin{aligned}
        \delta n (\boldsymbol{s}) &=\sum^{N_{\rm group}}_i \delta_{\rm D}(\boldsymbol{s}-\boldsymbol{s}_i)-\alpha\sum^{N_{\rm random}}_i \delta_{\rm D}(\boldsymbol{s}-\boldsymbol{s}_i)\,,
    \end{aligned}
\end{equation}
where $N_{\rm random}$ denotes the number of random points in the random catalog, and the factor $\alpha\equiv N_{\rm group}/N_{\rm random}=0.05$ in this work. %Here the fluctuation $\delta n$ is constructed so that its volume-average become zero, $\int d^3s \delta n (\boldsymbol{s})=0$, but $\delta T (\boldsymbol{s})$ does not.

\subsection{Estimator of the pairwise kSZ power spectrum multipoles}
\label{subsec:estimator}

We adopt the estimator of the density-weighted pairwise kSZ power spectrum multipoles developed by~\cite{sugiyama2018}, which is an analogy to the estimator of the galaxy power spectrum multipoles \citep{Feldman1994,Yamamoto05},
\begin{equation}
\label{eq:PkSZ_l_estimator}
    \begin{split}
        \hat{P}^{\ell}_{\rm kSZ}(\boldsymbol{k})=-\frac{2\ell+1}{A}\int d^3 s_1 \int d^3 s_2 e^{-i\boldsymbol{k}\cdot{} \boldsymbol{s}_{12}}\mathcal{L}_\ell(\hat{\boldsymbol{k}}\cdot{}\hat{\boldsymbol{n}}_{12})[\delta T (\boldsymbol{s}_1)\delta n(\boldsymbol{s}_2)-\delta n (\boldsymbol{s}_1)\delta T(\boldsymbol{s}_2)]\,,
    \end{split}
\end{equation}
where $\boldsymbol{s}_{12}=\boldsymbol{s}_1-\boldsymbol{s}_2$ and %$\mathcal{L}_\ell$ is the Legendre polynomial,  is the separation vector between $\boldsymbol{s}_1$ and $\boldsymbol{s}_2$, and 
the unit vector $\hat{\boldsymbol{n}}_{12}$ of $\boldsymbol{n}_{12}=(\boldsymbol{s}_1+\boldsymbol{s}_2)/2$ denotes the LOS direction to the mid-point of the pair of objects. These multipoles are normalized by~\citep{Feldman1994}
\begin{equation}
\label{eq:normalization}
    \begin{split}
        A=\int d^3 s\bar{n}^2(\boldsymbol{s})\,,\,\,\,\,\,\,\,\,\bar{n}(\boldsymbol{s})=\alpha \sum^{N_{\rm random}}_i \delta_{\rm D}(\boldsymbol{s}-\boldsymbol{s}_i)\,,
    \end{split}
\end{equation}
where the number density $\bar{n}$ is directly measured from the random catalog. We do not use the FKP weight in this estimator since it is not derived for an optimal kSZ detection.

The shot noise term disappears in Equation~(\ref{eq:PkSZ_l_estimator}), as it is canceled out by the subtraction of the two terms in the square bracket. This is also true for the detector noise and the primary CMB anisotropies contaminating the $\delta T(\boldsymbol{s})$ measurement. The pairwise estimator guarantees that they will not bias the kSZ power spectrum, although their contributions to the covariance matrix remain.

%The shotnoise term that often occurs in the galaxy multipoles (see \citep{Hand2017}) is cancelled out due to the difference between the the square braket of equation \ref{eq:Pkmu,l}. The even-parity noise, such as detector noise and CMB anisotropies, will not contribute to kSZ signal because the odd-pole Legendre polynomials in the pairwise estimator (equation \ref{eq:Pkmu,l}) \citep{sugiyama2018}.
Under the local plane-parallel approximation, %the Legendre polynomials are described as
\begin{equation}
\label{eq:plane_parallel}
    \begin{split}      \mathcal{L}_\ell(\hat{\boldsymbol{k}}\cdot{}\hat{\boldsymbol{n}}_{12})\approx\mathcal{L}_\ell(\hat{\boldsymbol{k}}\cdot{}\hat{\boldsymbol{s}}_{1})\approx\mathcal{L}_\ell(\hat{\boldsymbol{k}}\cdot{}\hat{\boldsymbol{s}}_{2})\,,
    \end{split}
\end{equation}
Equation~(\ref{eq:PkSZ_l_estimator}) can be simplified as \citep{Yamamoto05,sugiyama2018}
\begin{equation}
\label{eq:estimator}
    \begin{split}
        \hat{P}^{\ell}_{\rm kSZ}(\boldsymbol{k})=-\frac{2\ell+1}{A}[\delta T (\boldsymbol{k})\delta n ^* _\ell(\boldsymbol{k})-\delta T ^* (\boldsymbol{k})\delta n _\ell(\boldsymbol{k})]\,,
    \end{split}
\end{equation}
where
\begin{eqnarray}
 %   \begin{aligned}
        \delta n _\ell(\boldsymbol{k})&=&\int d^3 s e^{-i\boldsymbol{k}\cdot{} \boldsymbol{s}}\delta n (\boldsymbol{s})\mathcal{L}_\ell(\hat{\boldsymbol{k}}\cdot{}\hat{\boldsymbol{s}})\,,
        \label{eq:delta_n_l_k}\\
%    \end{aligned}\\
%    \begin{aligned}
        \delta T(\boldsymbol{k})&=&\int d^3 s e^{-i\boldsymbol{k}\cdot{} \boldsymbol{s}}\delta T(\boldsymbol{s})\,.
        \label{eq:delta_T_k}
%    \end{aligned}
\end{eqnarray}

The evaluation of the Legendre polynomial involved quantity ($\delta n_\ell$ here) is time consuming. It is proposed in \cite{Bianchi2015b,Scoccimarro2015} that, by decomposing $\hat{\boldsymbol{k}}\cdot\hat{\boldsymbol{s}}$ into its Cartesian components, Equation~(\ref{eq:delta_n_l_k}) can be expressed as a sum over the Fourier transforms of $\delta n(\boldsymbol{s})$ weighted by products of Cartesian vectors. Thus, the application of FFT technique\footnote{\href{https://numpy.org/doc/stable/reference/generated/numpy.fft.fftn.html}{numpy.fft.fftn}}  can speed up the calculation in several orders of magnitude as compared to a naive summation implementation of Equation~(\ref{eq:delta_n_l_k}). The number of FFTs required in each $\hat{P}_\ell$ calculation is $(\ell+1)(\ell+2)/2$ in this scenario. 

{Instead of the Cartesian decomposition, \cite{Hand2017} and~\cite{Sugiyama2018b} proposed to expand the Legendre polynomial into a product of spherical harmonics using  the spherical harmonic addition theorem,} 
\begin{equation}
\label{eq:sph_harm_decom}
    \begin{split}        \mathcal{L}_\ell(\hat{\boldsymbol{k}}\cdot{}\hat{\boldsymbol{s}})=\frac{4\pi}{2\ell+1}\sum^\ell_{m=-\ell}Y_{\ell m}(\hat{\boldsymbol{k}})Y^*_{\ell m}(\hat{\boldsymbol{s}})\,.
    \end{split}
\end{equation}
Then, Equation~(\ref{eq:delta_n_l_k}) becomes
\begin{equation}
 %   \begin{aligned}
        \delta n _\ell(\boldsymbol{k})=\frac{4\pi}{2\ell+1}\sum^\ell_{m=-\ell}Y_{\ell m}(\hat{\boldsymbol{k}})\int d^3 s e^{-i\boldsymbol{k}\cdot{} \boldsymbol{s}}\delta n (\boldsymbol{s})Y^*_{\ell m}(\hat{\boldsymbol{s}})\,.
        \label{eq:delta_n_l_k_Y}
\end{equation}
In this case, the implication of only $2\ell+1$ times of FFTs is needed to evaluate each $\delta n_\ell$. Although when $\ell=1$, both methods require 3 FFTs; the improvement will become evident as $\ell$ increases. Therefore, we adopt the second method to calculate the pairwise kSZ power spectrum dipole, and the calculation of Equation~(\ref{eq:delta_n_l_k_Y}) heavily refers to the software toolkit \texttt{nbodykit}~\citep{Hand2018}.

Finally, we select appropriate $k$ bins, $k_i\in [0.05,0.155]\ (h/\rm Mpc)$ with interval $\Delta k=0.01 h/\rm Mpc$, and average the $|\boldsymbol{k}|$ valves within each bin to obtain an effective $k$ value. This tackles the binning discreteness effect, which can be prominent at linear scales. Being dominated by noises and photo-z damping, larger $k$ modes cannot increase the measurement S/N. Furthermore, we average the measured $\hat{P}^\ell_{\rm kSZ}$ within each $k$ bin from Equation~(\ref{eq:estimator}) to compute our estimated $\hat{P}_{\rm kSZ}^\ell(k)$:
\begin{equation}
\label{eq:estimator_P}
    \begin{aligned}
        \hat{P}_{\rm kSZ}^\ell(k)&=\int\frac{d\Omega_k}{4\pi}\hat{P}^{\ell}_{\rm kSZ}(\boldsymbol{k})=\frac{1}{N_{\rm mode}}\sum_{k_i\le k<k_{i+1}}\hat{P}^{\ell}_{\rm kSZ}(\boldsymbol{k})
    \end{aligned}
\end{equation}
with
\begin{equation}
\label{eq:estimator_k}
    \begin{aligned}
        k=\frac{1}{N_{\rm mode}}\sum_{k_i\le k<k_{i+1}}|\boldsymbol{k}|\,,
    \end{aligned}
\end{equation}
where $N_{\rm mode}$ is the number of Fourier modes in a $k$ bin ($k_i\le k<k_{i+1}$).

\subsection{Measurement details}
\label{subsec:measurement}
Referring to \cite{Gilmarin2016a,sugiyama2018}, the power spectrum multipoles of NGC and SGC are weighted by their effective areas to obtain the combined power spectrum multipoles:
\begin{equation}
\label{eq:combined_P}
    \begin{aligned}
        \hat{P}_{\rm kSZ}^\ell(k)=\frac{A_{\rm NGC}\hat{P}_{\rm kSZ}^{\ell, \rm (NGC)}(k)+A_{\rm SGC}\hat{P}_{\rm kSZ}^{\ell, \rm (SGC)}(k)}{A_{\rm NGC}+A_{\rm SGC}}\,,
    \end{aligned}
\end{equation}
where $A_{\rm NGC}$ and $A_{\rm SGC}$ are effective coverage of the contiguous areas of NGC and SGC, namely $A_{\rm NGC}=9622\ \rm deg^2$ and $A_{\rm SGC}=8601\ \rm deg^2 $. Equation~(\ref{eq:combined_P}) is used analogically to estimate other parameters, such as the median redshift,  median mass of the whole group sample, etc.%In the following analysis, we carry out the parameter fitting process for the combined kSZ power spectrum of NGC and SGC.

The FFTs are performed on regular grids with a fiducial size of $512^3$. We establish the Cartesian coordinates $\boldsymbol{s}=(x,y,z)$ with the $z$-axis pointing to the north pole. Then, we locate the group samples into a cuboid of dimensions $(L_x,L_y,L_z)$, with $(L_x,L_y,L_z)[h^{-1}\rm Mpc]=(2400,4000,2600)$ for NGC, and $(2600,4000,3500)$ for SGC.  The averaged grid size is $\sim 6h^{-1}\rm Mpc$ for both NGC and SGC. The triangular-shaped cloud (TSC) assignment function is used to distribute both the groups and associated $\delta T$s onto the grids. This gridding process induces numerical artifacts like the aliasing effect and the window function effect (e.g., \cite{Jing05}). In Fourier space, we follow the interlacing technique proposed by \cite{Hockney81,Sefusatti2016} to alleviate the aliasing effect, and correct the window function (e.g., Equation~(18) of \cite{Jing05}) for each $\boldsymbol{k}$ mode on the grids.  As illustrated by figure~2 of \cite{Hand2018}, these corrections guarantee to reduce the corresponding artifacts down to a subpercent level even at scales near the Nyquist frequency in our case and will not bias the results significantly. We also try the measurement on a $1024^3$ grid, and little difference can be observed within the $k$ range we use. More details about the FFT related procedure can be found in Appendix B2 of \cite{sugiyama2018}.

\subsection{Covariance matrix estimation}
\label{sec:Cij}
We perform the delete-one jackknife (JK) method \citep{sugiyama2018} to estimate the data covariance through resampling of the date itself. We employ the {\it kmeans} algorithm$\footnote{\href{https://github.com/esheldon/kmeans_radec/}{https://github.com/esheldon/kmeans\_radec/}}$ on the random catalog to divide the group samples into $N_{\rm JK}$ spatial subregions on the sky~\citep{Kwan2017}. %To ensure that each sub-region cover approximately the same area, we run kmean sample algorithm with the random and denser random catalogue to find sub-region centers. 
We employ $N_{\rm JK}=200$, in which $N_{\rm JK}^{(\rm NGC)}=105$, and $N_{\rm JK}^{(\rm SGC)}=95$ so that their ratio is close to corresponding sky area ratio.

The covariance matrix in the JK scheme is estimated by
\begin{equation}
\label{eq:CJK}
    \begin{aligned}
        \hat{C}_{\rm JK,ij}=\frac{N_{\rm JK}-1}{N_{\rm JK}}\sum^{N_{\rm JK}}_{a=1} (y_i^a-\bar{y}_i)(y_j^a-\bar{y}_j),
    \end{aligned}
\end{equation}
where $y_i^a=\hat{P}^{\ell=1,a}_{\rm kSZ}(k_i)$ is the power spectrum dipole at $k_i$ of the $a$th resampling data configuration, and its expectation value is 
\begin{equation}
\label{eq:mean yi}
    \begin{aligned}
        \bar{y}_i=\frac{1}{N_{\rm JK}}\sum^{N_{\rm JK}}_{a=1} y_i^a.
    \end{aligned}
\end{equation}

\cite{Hartlap2007} rescales the inverse covariance matrix $C_{\rm JK}^{-1}$ as 
\begin{equation}
\label{eq:C-1}
    \begin{aligned}
        \hat{C}^{-1}=\frac{N_{\rm JK}-N_{\rm bin}-2}{N_{\rm JK}-1}\hat{C}_{\rm JK}^{-1}
    \end{aligned}
\end{equation}
to get an unbiased precision matrix $\hat{C}^{-1}$. Although the original derivation of the Hartlap factor is not based on the JK technique, we apply this factor here to calculate a conservative precision matrix for our parameter inference procedure.

\subsection{Survey window functions}
\label{subsec:window}
%=============================================fig
\begin{figure}[t!]
\centering
\includegraphics[width=1.0\textwidth]{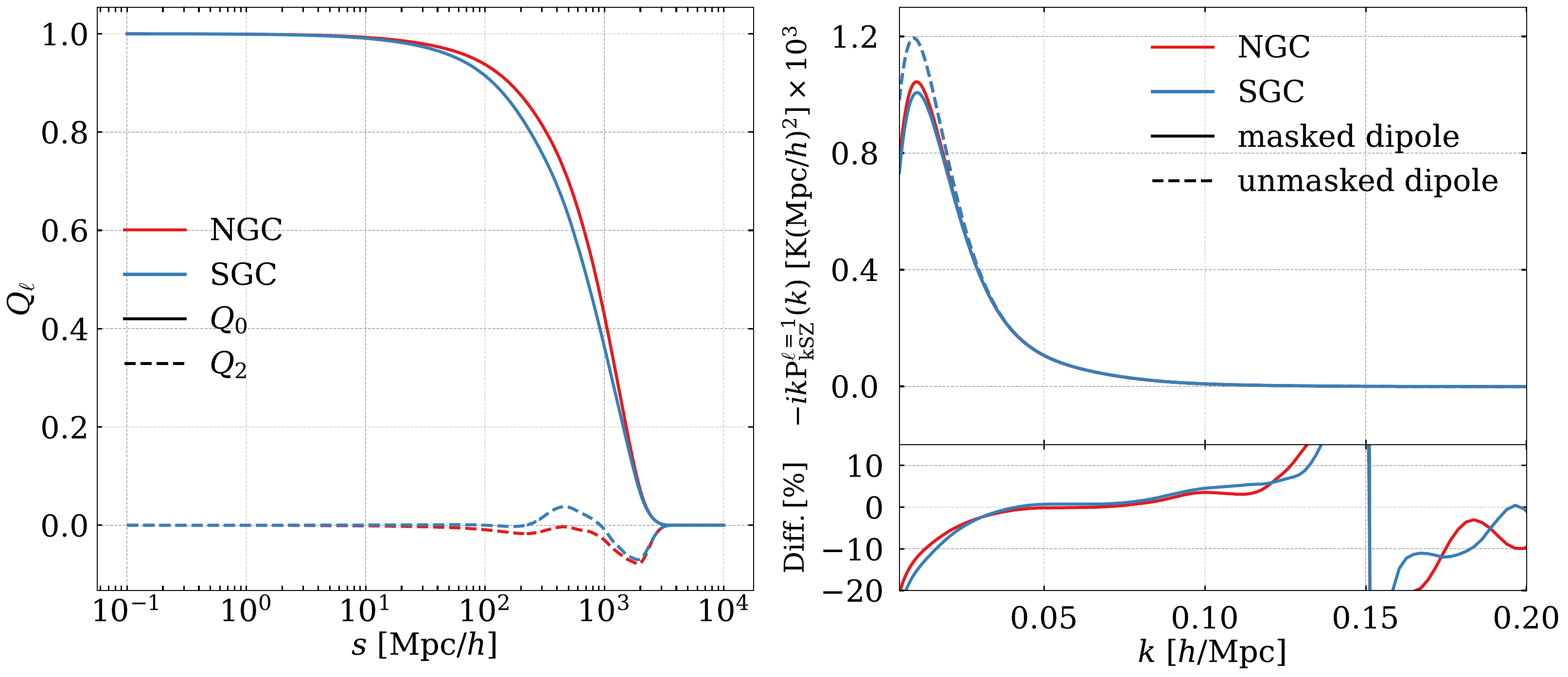}
\caption{\label{fig:wf_kSZ} {\textit{Left:} Multipoles of the survey window function for the baseline sample, which are used to compute the masked theoretical pairwise kSZ power dipole. Solid and dashed lines indicate the monopole and qradrupole separately. The red and blue lines represent the window functions of NGC and SGC respectively. \textit{Right:} Masked (solid lines) and unmasked (dashed lines) theoretical predictions of the pairwise kSZ dipole for the baseline sample. The red and blue lines are for NGC and SGC respectively. On the lower panel, the relative difference ${\rm Diff} = P_{\rm mask}/P_{\rm nomask}-1$ induced by the window function effect is presented. The inflation of ``Diff'' at $k\sim 0.15h/\rm Mpc$ is due to the zero-crossing of $P^{\ell=1}_{\rm kSZ}$ at that scale.}}
\end{figure}
%=============================================fig
% %=============================================fig
% \begin{figure}[t!]
% \centering
% \includegraphics[width=0.6\textwidth]{./maskedPkSZ.pdf}
% \caption{\label{fig:maskPkSZ} Masked (solid lines) and unmasked (dashed lines) theoretical predictions of the pairwise kSZ dipole for the baseline sample. The red and blue lines are for NGC and SGC respectively. On the lower panel, the relative difference ${\rm Diff} = P_{\rm mask}/P_{\rm nomask}-1$ induced by the window function effect is of the order of $10-20\%$ around $k\sim 0.01h/\rm Mpc$. {The inflation of `Diff' at $k\sim 0.15h/\rm Mpc$ is due to the zero-crossing of $P^{\ell=1}_{\rm kSZ}$ at that scale.}} %\LSH{The theoretical dipole has an insignificant positive and negative reversal at $k\sim 0.12h/\rm Mpc$ due to the FoG effect.  delete???}}
% \end{figure}
% %=============================================fig
In this paper, we follow the treatment suggested by \cite{Wilson2017,Beutler2017} to include the window function effect caused by the masked survey sky in the kSZ power spectrum model.  In this methodology, the window function is first calculated in configuration space and multiplies the pairwise kSZ correlation function to give the masked correlation function. This  masked correlation function is then Hankel transformed to produce the masked pairwise kSZ power spectrum.

The window function multipoles characterizing the distortion of the survey geometry are calculated by the random catalog and are given by
\begin{equation}
\label{eq:Ql}
    \begin{aligned}
        Q_{\ell}(s)\propto\sum_{\boldsymbol{s}_1}\sum_{\boldsymbol{s}_2}\frac{1}{s^3}RR(s,\hat{\boldsymbol{n}}_{12}\cdot{}\hat{\boldsymbol{s}})\mathcal{L}_\ell(\hat{\boldsymbol{n}}_{12}\cdot{}\hat{\boldsymbol{s}})\delta_D(s-|\boldsymbol{s}_{12}|),
    \end{aligned}
\end{equation}
where $RR$ is the number of random pairs in a logarithmic $s$ bin$\footnote{\href{https://corrfunc.readthedocs.io}{Corrfunc.mocks.DDsmu\_mocks}}$. Here, $Q_\ell$ is normalized to enforce $Q_0(0)=1$. %, and notice that $s$ is binned in logarithmic space. 
{In order to speed up the calculation, for samples with $ M_{\rm th}\ge 10^{13}M_\odot/h$, we randomly dilute the random catalog to the size of the $ M_{\rm th}= 10^{13}M_\odot/h$ group sample, and for samples with $M_{\rm th}<10^{13}M_\odot/h$, we dilute the random catalog to the size of the corresponding group sample size.} A convergence test in Appendix~\ref{app:window} shows that this dilution induces negligible systematic effects on the calculation of $\tilde{P}_{\rm kSZ}^{\ell=1}$ in Equation~(\ref{eq:PkSZ1}). 
%don't use the whole random catalogue, but randomly select one fifth or one tenth of them to calculate the $RR$. This choice is based on the random catalogue which has no other information except the geometric structure. And 

The left panel of figure~\ref{fig:wf_kSZ} shows the multipole components of the survey window function applied in our analysis. We expect that monopole and quadrupole respectively tend to unit and zero when $s\rightarrow{}0$, and both of them are equal to zero in dimensions beyond the survey volume. The positive quadrupole values reflect that, for a fixed $s$ bin, there are more random pairs found along the LOS than perpendicular to the LOS, and vice versa.% {The positive and negative values of quadrupole reflect the volume ratio of parallel and vertical lines of sight, in which the positive quadurpole means the volume is bigger in the parallel direction. }

%The masked pairwise kSZ power spectrum on the observation is represented by Equation~(\ref{eq:estimator_P}). 
In turn, the masked kSZ power spectrum including the window function can be theoretically calculated by~\citep{sugiyama2018}
\begin{equation}
\label{eq:Pkl}
    \begin{aligned}
        \tilde{P}_{\rm kSZ}^\ell(k)=4\pi(-i)^\ell(2\ell+1)\int ds\ s^2 j_\ell(ks)\sum_{\ell_1\ell_2}\left(
        \begin{array}{c}
            \ell\ \ell_1\ \ell_2 \\
            0\ 0\ 0 
        \end{array}
        \right)^2\xi_{\rm kSZ}^{\ell_1}(s)Q_{\ell_2}(s),
    \end{aligned}
\end{equation}
where $\xi_{\rm kSZ}^{\ell}(s)$ is given by
\begin{equation}
\label{eq:theoretical xi,l}
   \begin{split}
       \xi_{\rm kSZ}^{\ell}(s)=i^\ell \int\frac{k^2dk}{2\pi^2}j_\ell(ks)P_{\rm kSZ}^{\ell}(k).
   \end{split}
\end{equation}
Here, $j_\ell$ is the spherical Bessel function of order $\ell$. For the dipole, the expansion of Equation~(\ref{eq:Pkl}) can be linearly truncated as
\begin{equation}
\label{eq:PkSZ1}
    \begin{aligned}
        \tilde{P}_{\rm kSZ}^{\ell=1}(k)=-i4\pi\int ds\ s^2 j_{\ell=1}(ks)\xi_{\rm kSZ}^{\ell=1}(s)\left(Q_{0}(s)+\frac{2}{5}Q_{2}(s)\right),
    \end{aligned}
\end{equation}
in which the contributions of higher multipoles of the kSZ correlation function (e.g., the octopole $\xi_{\rm kSZ}^{\ell=3}$) %in the expansion of the pairwise kSZ power dipole 
are small and can be ignored. The robustness of this linear truncation is verified in Appendix~\ref{app:high_ell_expansion}. %\LSH{in which the contribution of higher multipoles of the kSZ correlation function is so small and can be ignore, since higher correlation function multipoles (e.g. the octopole $\xi_{\rm kSZ}^{\ell=3}$) and higheor window function multipoles (e.g. $Q_{\ell\geq2}$) are small, simultaneously. Relative test is shown in appendix~\ref{app:high_ell_expansion}.} 
Equation~(\ref{eq:PkSZ1}) is the final expression used to fit the measurement of the pairwise kSZ power dipole in this work. Its detailed derivation is presented in Appendix C of~\cite{sugiyama2018}.%, and for the convenience of readers, we repeat it in appendix~\ref{app:window}.

The theoretical predictions of the pairwise kSZ dipoles with and without the survey window functions are shown on the right panel of figure~\ref{fig:wf_kSZ}. The amplitude of the masked $P^{\ell=1}_{\rm kSZ}$ becomes smaller than the true one at large-scales, due to the lack of group pairs beyond the survey volume. The window function effect becomes significant at $k<0.05h/\rm Mpc$ and is of the order of $10\%-20\%$ around $k\sim0.01h/\rm Mpc$. Since our measurement S/N is dominated by linear scales, this effect is a key systematic error that ought to be corrected in our analysis. %\LSH{The theoretical dipole has an insignificant positive and negative reversal at $k\sim 0.12h/\rm Mpc$ due to the FoG effect.  delete???}

{
\subsection{Imaging systematics}
\label{subsec:imaging_sys}
%=============================================fig
\begin{figure}[t!]
\centering
\includegraphics[width=1.0\textwidth]{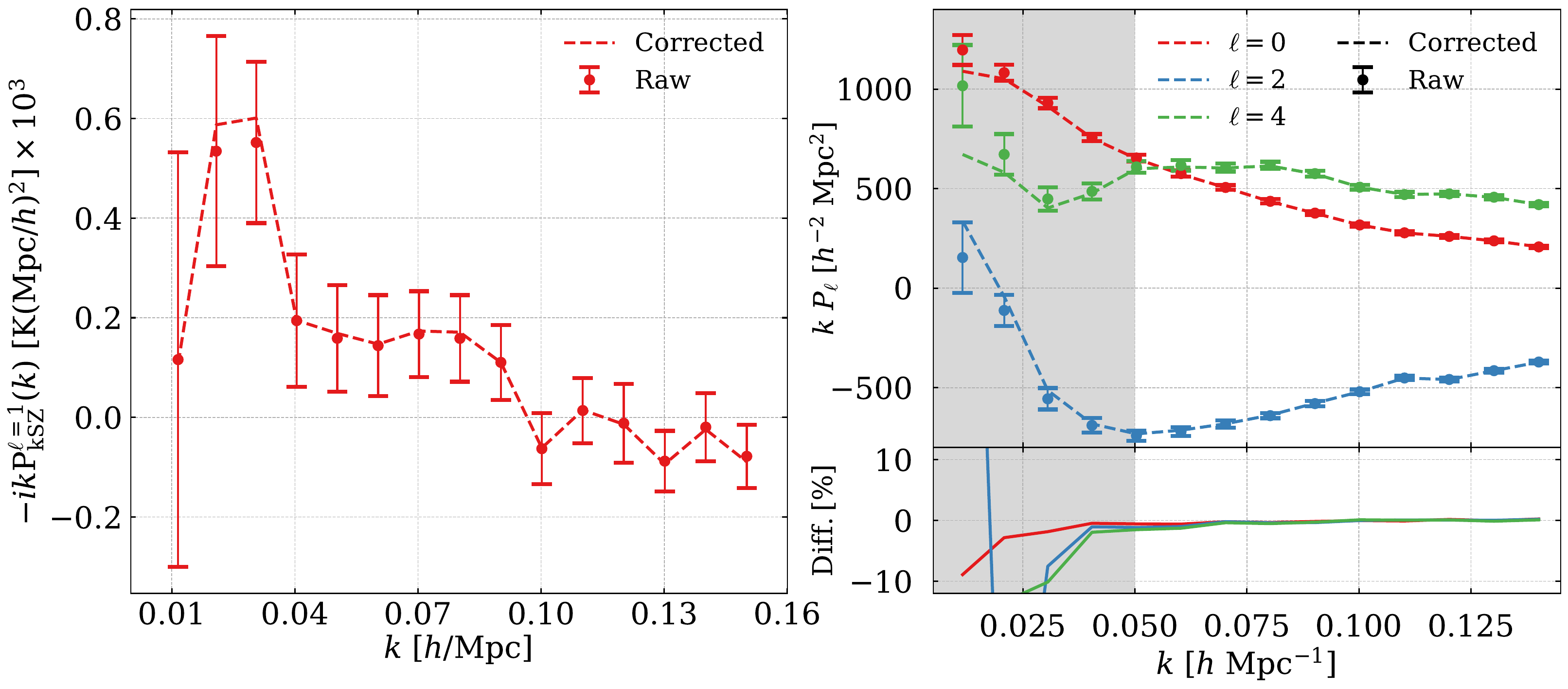}
\caption{\label{fig:imaging_sys_test} {{\it Left:} the pairwise kSZ dipole measurement of the baseline sample without (data points with error bars) and with (dashed line) the imaging-systematics correction. {\it Right:} The multipoles of redshift space density power spectrum without (data points with error bars) and with (dashed line) the imaging-systematics correction. The lower panel shows the fractional difference between those of ``corrected'' and ``raw'' ones. Measurements are made from the baseline sample.}}
\end{figure}
%=============================================fig
It has been well demonstrated that a variety of observation conditions, such as stellar contamination, Galactic extinction, sky brightness, seeing, and airmass, can introduce spurious fluctuations in the measured galaxy density field~\citep[e.g.,][]{Scranton2002a, Myers2006, Morrison2015, Kitanidis2019}. These fluctuations, known as imaging systematics, must be addressed in order to obtain reliable results for large-scale clustering studies. The group catalogs used in this work are derived from photometric galaxy catalogs, which may be subject to such biases. %After examining the group surface density as a function of imaging conditions and cross-correlating the groups with imaging maps, we conclude that our sample is affected by imaging systematics. \textbf{need to show plots? I have the plots.}

A widely used approach to reduce the imaging systematics is to assign weights to the targets so that the weighted sample does not show any correlation with the various observation conditions~\citep[e.g.][]{Ross2017, Bautista2018}. This can be achieved by minimizing the difference between the predicted and observed target surface density. The predicted target density is simply a function of various imaging maps, which are designed to not contain any cosmological signal. Previous studies assumed a linear functional form~\citep[e.g.,][]{Ross2011a}, which may not be able to capture the nonlinear dependence on imaging systematics in areas with strong contamination, such as close to the Galactic plane~\citep[e.g.,][]{Ho2012}. Recently, a machine-learning-based approach has been used to capture these complex relationships~\cite[e.g.,][]{Rezaie2020, Chaussidon2021}.

We reduce the imaging systematics in our targets using the random forest algorithm, which is available in the
GitHub code {\sc regressis}\footnote{\url{https://github.com/echaussidon/regressis}} developed by \citet{Chaussidon2021}. \citet{Xu2023} adapted the code to remove imaging systematics in autocorrelations and cross-correlations between multiple tomography bins in the galaxy samples from DR9 of the DESI Legacy Imaging Survey. Here, we follow the steps outlined in \citet{Xu2023} to mitigate imaging systematics and refer interested readers to their paper for details.

The comparison between the measured $\hat{P}_{\rm kSZ}^\ell$ with and without correcting imaging systematics is shown on the left panel of figure~\ref{fig:imaging_sys_test}. We see little systematic offsets when we ignore the imaging systematics. On the right panel of figure~\ref{fig:imaging_sys_test}, we also find little differences between redshift space density power spectrum multipoles $\hat{P}_\ell$ with and without the imaging-systematics correction, in which $\hat{P}_\ell$ is analyzed in Appendix~\ref{app:RSD_analysis} to constrain free parameters and determine the theoretical $P_{\rm pv}$ template in the $\bar{\tau}$ fitting. As a result, we consider the imaging systematics as a negligible systematic error source in our analysis, and we will not correct them in this work.
}

\section{Results}
\label{sec:result}

% %=============================================fig
% \begin{figure}[t!]
% \centering
% \includegraphics[width=0.6\textwidth]{./cij.pdf}
% \caption{\label{fig:cij}Correlation coefficients of the measured  pairwise kSZ power dipole from the baseline sample using the JK method by applying $\theta_{\rm AP}=4.2^\prime$.}
% \end{figure}
% %=============================================fig

%The $\rm S/N$ and the optical depth $\tau$ with $\theta_{\rm AP}$ as the variable measured from the Planck map with the richest-200000-groups samples are shown in figure \ref{fig:SN}.

In this section, we presents main results of this paper, including the pairwise kSZ dipoles, the fitted $\bar{\tau}$ values, and the $\bar{\tau}-\tilde{M}$ relationship.

\subsection{Fitting procedure}
\label{subsec:fit_procedure}

%\LSH{Referring to~\citep{Chen2022}}, 
We estimate the average optical depth $\bar{\tau}$ of each group sample by fitting the measured kSZ dipole  with the template given by Equation~(\ref{eq:PkSZ1}). %As illustrated in appendix~\ref{app:NL-kSZ}, we have set $\sigma_v=0\ km/s$ in the model, and 
$\bar{\tau}$ is the only free parameter to be fitted. The $\chi^2$ of the fitting is %and it also exists in the template as a proportionality constant so that we perform a concise $\chi^2$ minimization with the covariance matrix derived in Section \ref{sec:Cij} to find the best-fitting parameter. The $\chi^2$ is given by \citep{Chen2022}
\begin{equation}
\label{eq:chi2}
    \begin{aligned}
        \chi^2(\bar{\tau})=[\hat{P}-\bar{P}(\bar{\tau})]^{\rm T} \hat{C}^{-1}[\hat{P}-\bar{P}(\bar{\tau})],
    \end{aligned}
\end{equation}
where $\hat{C}^{-1}$ is the precision matrix, $\hat{P}$ is the measured kSZ dipole with the vector size $N_{\rm bin}=15$, and $\bar{P}$ is the theoretical with $\bar{\tau}$ as the single variable. 

The best-fitted value and the associated statistical error of $\bar{\tau}$ in this linear fitting can be analytically given by~\citep{Chen2022} 
\begin{equation}
\label{eq:tau}
    \begin{aligned}
        \bar{\tau}_{\rm bestfit}=\frac{\hat{P}^{\rm T} \hat{C}^{-1}\bar{P}(\bar{\tau}=1)}{\bar{P}(\bar{\tau}=1)^{\rm T}\hat{C}^{-1}\bar{P}(\bar{\tau}=1)}\,,
    \end{aligned}
\end{equation}

\begin{equation}
\label{eq:tau_err}
    \begin{aligned}
            \sigma^2_{\bar{\tau}} = \frac{1}{\bar{P}(\bar{\tau}=1)^{\rm T}\hat{C}^{-1}\bar{P}(\bar{\tau}=1)}\,.
    \end{aligned}
\end{equation}

Finally, we calculate the S/N of the measurement by
\begin{equation}
\label{eq:SN}
    \begin{aligned}
        \frac{\rm S}{\rm N}=\sqrt{\chi^2_{\rm null}-\chi^2_{\rm min}},
    \end{aligned}
\end{equation}
where $\chi^2_{\rm null}=\chi^2(\bar{\tau}=0)$ and $\chi^2_{\rm min}=\chi^2(\bar{\tau}=\bar{\tau}_{\rm bestfit})$

\subsection{Pairwise kSZ dipole and $\bar{\tau}$}
\label{subsec:result_dipole}
%=============================================fig
\begin{figure}[t!]
\centering
\includegraphics[width=0.46\textwidth]{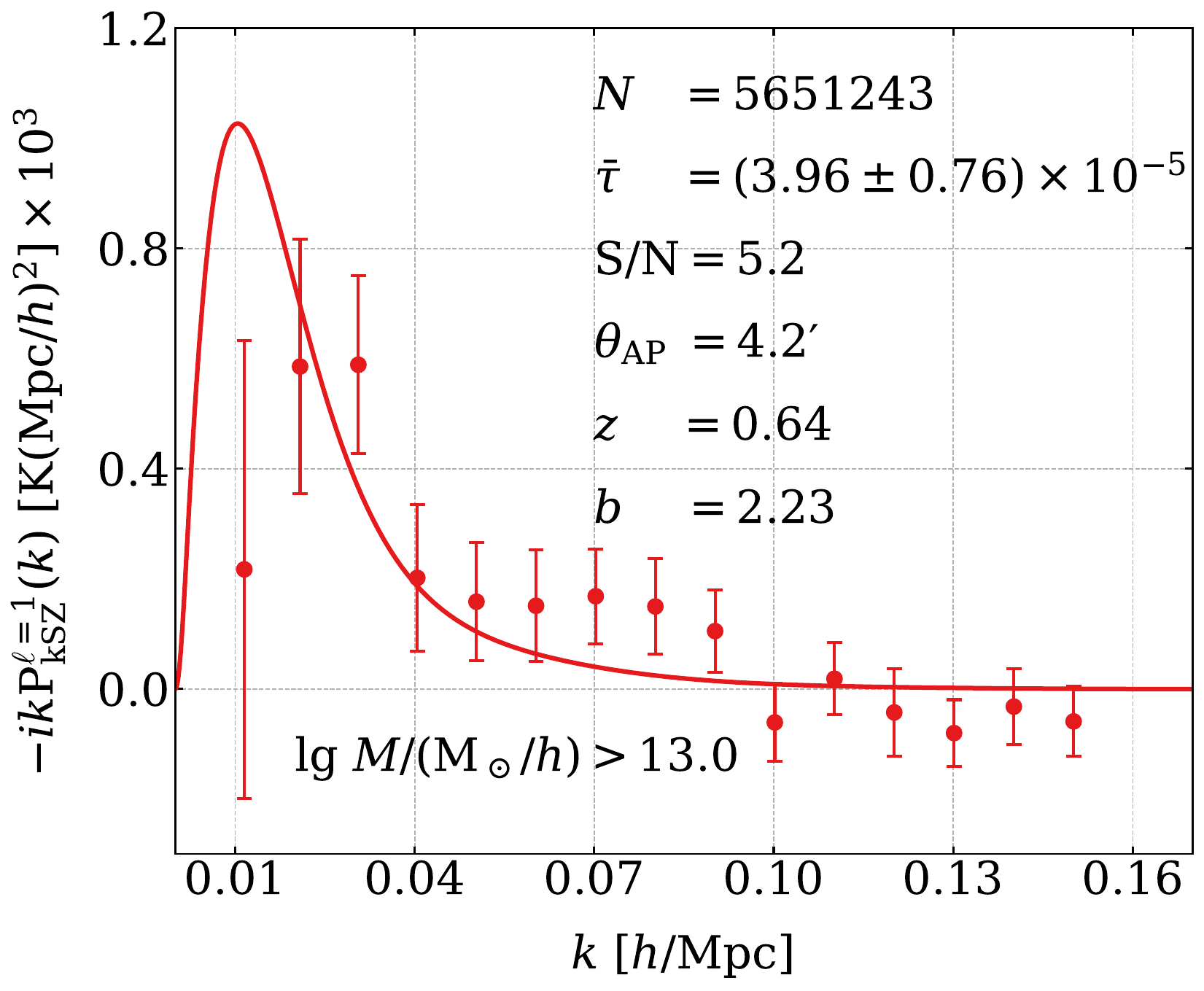}
\includegraphics[width=0.50\textwidth]{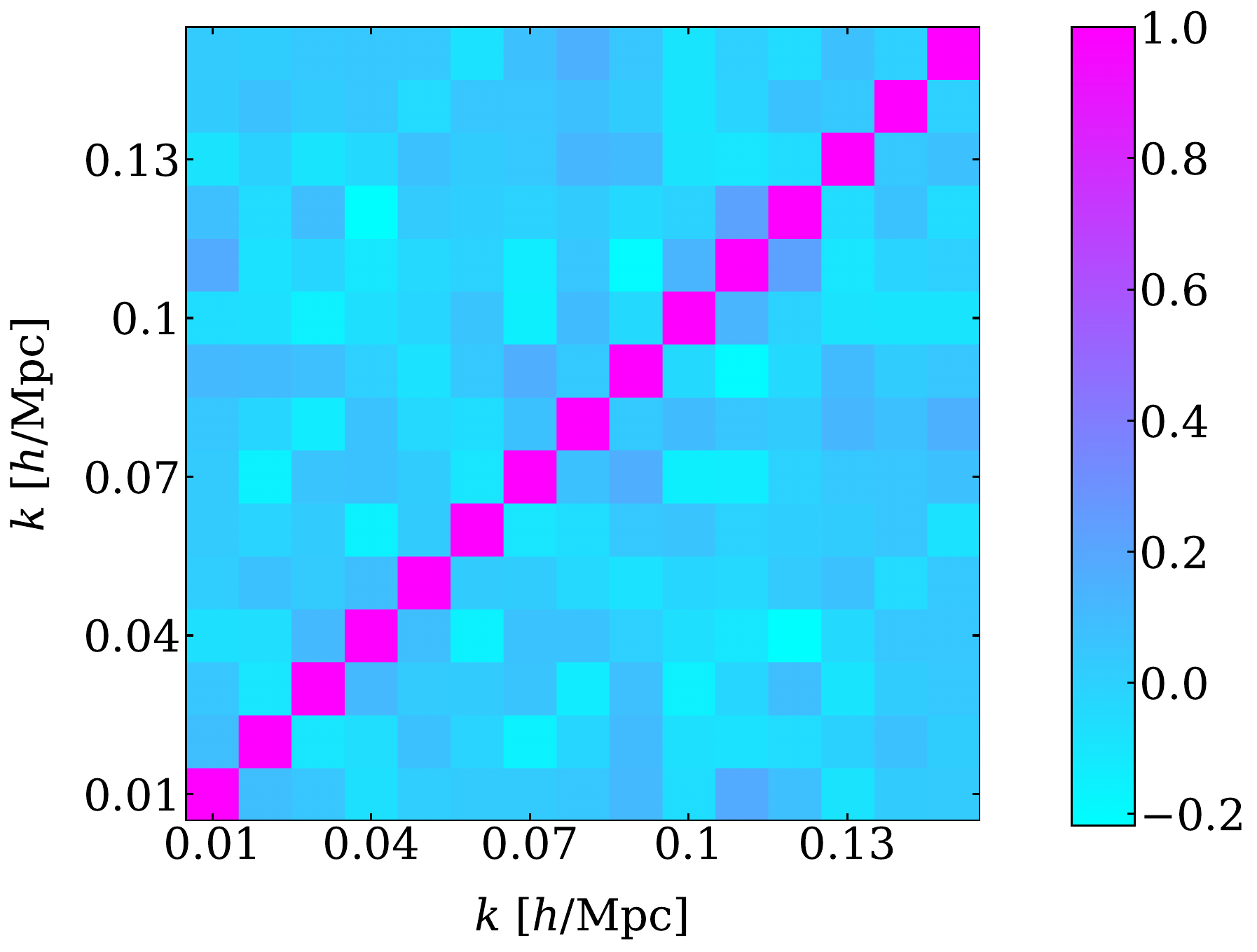}
\caption{\label{fig:Pkl_kSZ_detection} {\textit{Left:} The pairwise kSZ dipole measurement of the baseline sample compared with the theoretical template (solid line). By applying $\theta_{\rm AP}=4.2^\prime$, the best-fitted $\bar{\tau}=(3.96\pm0.76)\times10^{-5}$ with the measurement $\rm S/N=5.2$. \textit{Right:} correlation coefficients of the measured  pairwise kSZ power dipole from the baseline sample using the JK method by applying $\theta_{\rm AP}=4.2^\prime$.}}
\end{figure}
%=============================================fig
%=============================================fig
\begin{figure}[t!]
\centering
\includegraphics[width=1\textwidth]{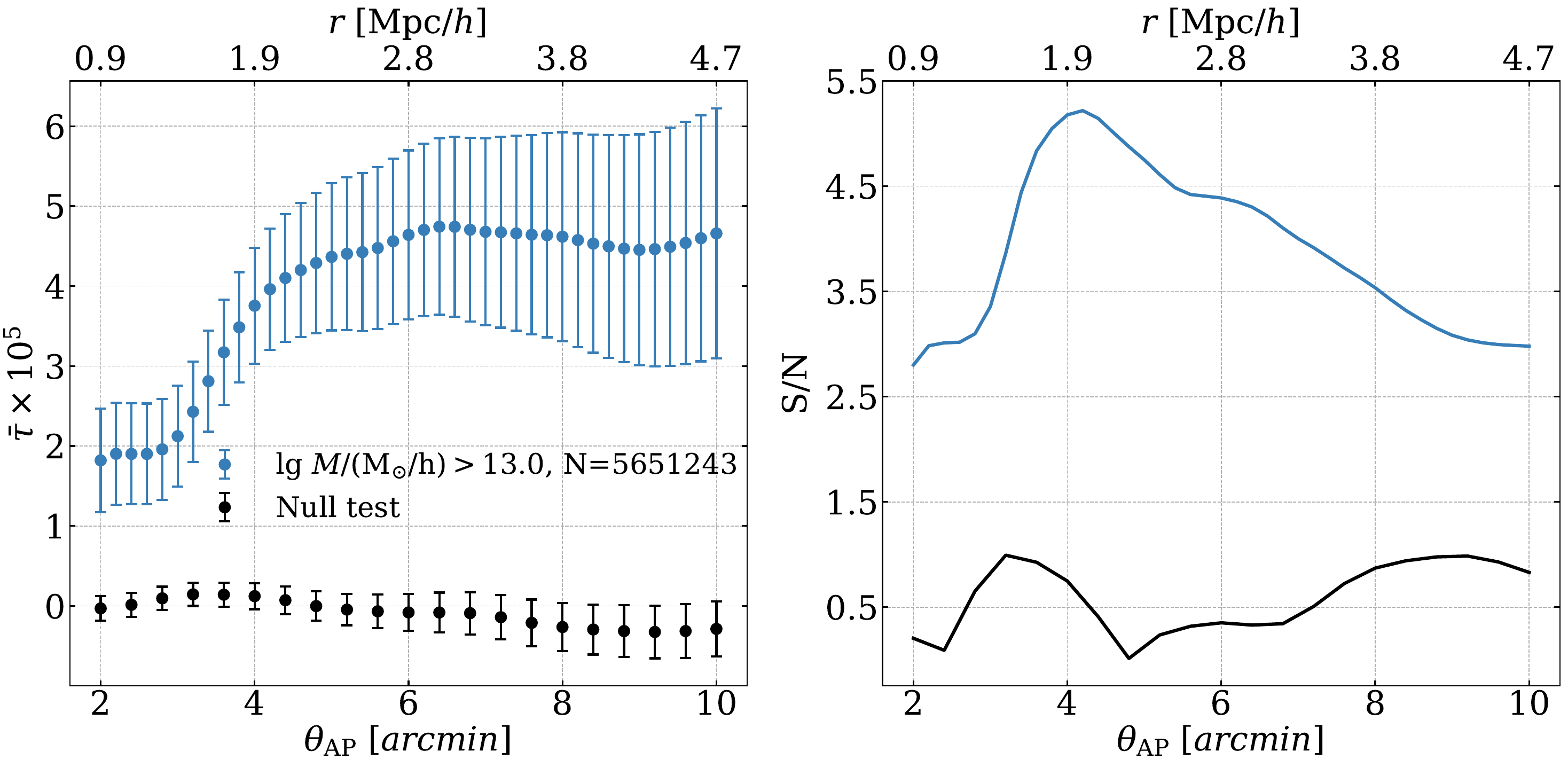}
\caption{\label{fig:tau_profile} {{\it Left:} Measured $\bar{\tau}$ profiles of the baseline sample by choosing different AP filter radii ($\theta_{\rm AP}$). The blue dots with error bars represent the $\bar{\tau}$ profile of the baseline sample, and the black dots with error bars show the results of the null test by targeting a random sample having the same size of the baseline sample. On top of the figure, we label the comoving radius corresponding to $\theta_{\rm AP}$. {\it Right:} corresponding detection $\rm S/N$s. }}% are shown with fixed $\theta_{\rm AP}$ and non-fixed $\theta_{\rm AP}$ ($\beta$), respectively, as the variables on the upper panel. And the corresponding optical depth $\tau$s with error bars are on the lower panel.}
\end{figure}
%=============================================fig
%$N_{\rm mem}$ is the main criteria of our sample selection. 
In Appendix~\ref{app:diff_sample}, we vary the lower mass threshold to select the heaviest $1.6\times10^7$ to $1.8\times10^5$ groups. %We also vary $M$\&$\theta_{\rm halo}$ to select clusters from $M>10^{13.4}M_\odot/h$ to $M>10^{14.2}M_\odot/h$ and from $\theta_{\rm halo}>2.0^\prime$ to $\theta_{\rm halo}>4.0^\prime$. 
For each sample, we change $\theta_{\rm AP}$  and see how the measurement S/N evolves with it (figures~\ref{fig:SN_all}). We find that the baseline sample, namely the heaviest $\sim 5.65\times10^6$ groups, has the highest S/N when $\theta_{\rm AP}=4.2'$. We present its dipoles and associated results here.

The left panel of figure~\ref{fig:Pkl_kSZ_detection} is the main result of this work, which shows {the highest ${\rm S/N}=5.2$ dipole measurement} from the baseline sample, with $\theta_{\rm AP}=4.2'$ and its covariance. This serves as one of the highest S/N kSZ detections in the literature, and is the highest S/N kSZ detection in Fourier space currently. As expected, the useful kSZ signal dominates only at linear scales since it is diluted by the photo-z damping and noises (shot noise, CMB detector noise, and primary CMB contamination) at nonlinear scales. The nonlinear kSZ model (solid lines) gives a good fit to the data, and the {fitted $\bar{\tau}=(3.96\pm0.76)\times10^{-5}$ }for $\tilde{M}=1.75\times10^{13}M_\odot/h$. On the right panel of figure~\ref{fig:Pkl_kSZ_detection}  we present the correlation coefficients of the dipole measurements from baseline sample. The off-diagonal elements are subdominant in the covariance matrix, and the kSZ field on each $k$ scale evolves independently with each other.

By varying $\theta_{\rm AP}$, we display in figure~\ref{fig:tau_profile} the measured $\bar{\tau}$ profile of the baseline sample. This profile encodes valuable information about the gas distribution within halos, and it can be fitted by some gas density profile model to constrain the gas fraction $f_{\rm gas}$ within halos~\citep{Amodeo2021,sugiyama2018}. In a companion paper, we will study this $\bar{\tau}$ profile with adequate profile models, e.g., the baryonification models~\citep{Schneider2015,Giri2021,Arico2021}, and constrain the halo gas content in our samples.

In addition, we make a null test by targeting $N$ random points within the DESI sky region, where $N$ is the group number of the base line sample. Going through the same pipeline, the measured $\bar{\tau}$ profile of this random sample is consistent with zero, as shown in figure~\ref{fig:tau_profile}. This null test further validates the robustness of our kSZ detections.

\subsection{$\bar{\tau}-\tilde{M}$ relationship}
\label{subsec:tau_M}
% %=============================================fig
% \begin{figure}[t!]
% \centering
% \includegraphics[width=0.48\textwidth]{./tau_mass.pdf}
% \includegraphics[width=0.48\textwidth]{./bias1tau_mass.pdf}
% \caption{\label{fig:tau-mass} {\textit{Left:} The measured $\bar{\tau}-\tilde{M}$ relationship with $\theta_{\rm AP} = 4.2'$. The points with error bars are measurements from various samples with different $M_{\rm th}$ and the solid lines are the best-fitted Equation~(\ref{eq:tau_mass}) with $\beta=(0.98^{+0.15}_{-0.15})\times10^{-4}$ and $\gamma=0.55^{+0.10}_{-0.10}$. We caution that the error bars of different $\bar{\tau}$s are correlated with each other, so the fitted error bars for $\beta$ and $\gamma$ are underestimated here. \textit{Right:} The $b_1\bar{\tau}-\tilde{M}$ relationship calculated by multiplying the $\bar{\tau}-\tilde{M}$ relation on the left panel by the best-fitted $b_1$ values in appendix~\ref{app:RSD_analysis}.}}
% \end{figure}
% %=============================================fig
%=============================================fig
\begin{figure}[t!]
\centering
\includegraphics[width=0.6\textwidth]{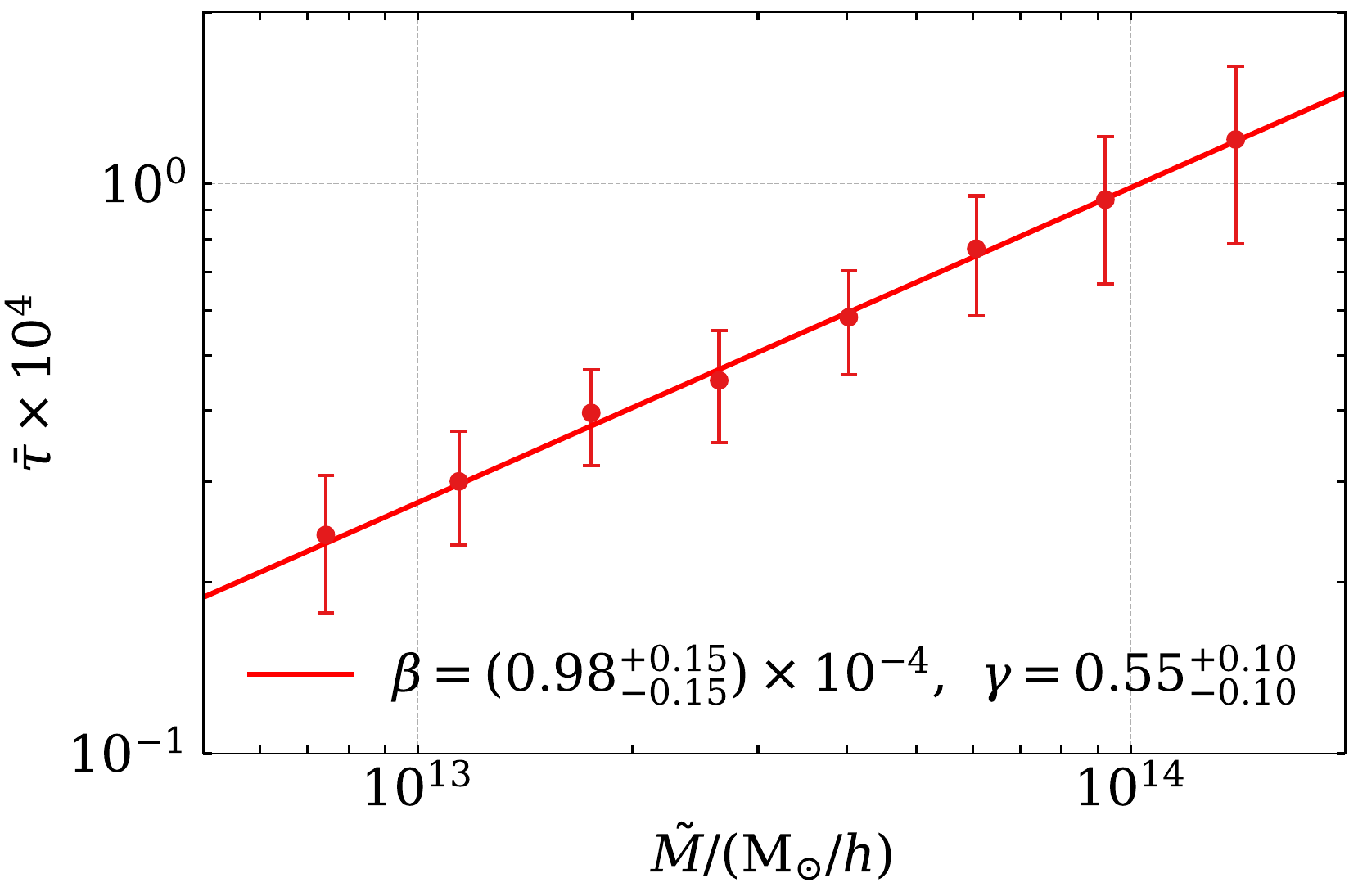}
\caption{\label{fig:tau-mass} {The measured $\bar{\tau}-\tilde{M}$ relationship with $\theta_{\rm AP} = 4.2'$. The points with error bars are measurements from various samples with different $M_{\rm th}$ and the solid lines are the best-fitted Equation~(\ref{eq:tau_mass}) with $\beta=(0.98^{+0.15}_{-0.15})\times10^{-4}$ and $\gamma=0.55^{+0.10}_{-0.10}$. We caution that cross-correlations between different $\bar{\tau}$s are ignored in this fitting, so the fitted error bars for $\beta$ and $\gamma$ are underestimated here.}}
\end{figure}
%=============================================fig
Next, we explore the relationship between the average optical depth and the median halo mass of a sample. We choose the measured optical depths of different samples in appendix~\ref{app:diff_sample} measured with $\theta_{\rm AP}=4.2'$ and fit the $\bar{\tau}-\tilde{M}$ relationship with a simple power law, 
\begin{equation}
\label{eq:tau_mass}
    \begin{aligned}
%        \tau= \beta (\frac{M}{10^{14}\rm M_\odot/h})^\gamma\,,
        \log\bar{\tau} = \gamma(\log \tilde{M}-14)+\log\beta
    \end{aligned}
\end{equation}
where $\beta$ and $\gamma$ are two free parameters. 

{The results are shown in figure~\ref{fig:tau-mass}. The red data points with error bars are measured $\bar{\tau}$s associated with the median group sample masses, and the red solid line is the best-fitted Equation~(\ref{eq:tau_mass}). The fitted  $\gamma=0.55^{+0.1}_{-0.1}$ is roughly consistent with the $\bar{\tau}\propto \tilde{M}$ relation, as expected for massive groups, and this result confirms our motivation of sample selection with the group mass. In~\cite{Chen2022}, a higher value of $\gamma=1.25\pm0.14$ is found. Although being deviated from each other, the discrepancy between two results is not as large as it looks here. Since group samples in both works are not independent of each other by definition, the uncertainty of $\gamma$ is moderately underestimated in both papers. Albeit, with this, the disagreement can come from several factors. (1) A minor one is that two works use different AP filter radii, with $\theta_{\rm AP}=3.4'$ in~\cite{Chen2022}, and $\theta_{\rm AP}=4.2'$ here. (2) In~\cite{Chen2022}, most group samples used to fit Equation~(\ref{eq:tau_mass}) are heavy clusters with masses larger than $10^{14}{\rm M}_{\odot}/h$, while in our work the mass range of group samples is ranging from $10^{13}{\rm M}_{\odot}/h$ to $10^{14}{\rm M}_{\odot}/h$. (3) The redshift distribution of the group sample is affected by the mass threshold, so the sample redshift distributions are not identical to each other, and can introduce potential systematic biases in the measured $\bar{\tau}-\tilde{M}$ relationship. (4) Two works treat the density biases ($b_1$, $b_2$) and the photo-z uncertainty $\sigma_{\rm pho-z}$ in different manners. In this work, we apply a joint analysis of kSZ and RSD effects on the same data set. We first constrain $b_1$, $b_2$ and $\sigma_{\rm pho-z}$ from the measured $\hat{P}_\ell$, these determined free parameters are then adopted to determine $P_{\rm pv}$ and fit $\bar{\tau}$. In~\cite{Chen2022}, they evaluated $b_1$ from the \textit{Sheth01} $b_1-M$ model~\citep{Sheth2001}. Then, $\xi_{\rm pv}$ was directly measured from the simulation halo catalog constructed to match the calculated $b_1$, on which the photo-z uncertainties are directly applied on each halo. }

%In all these possibilities, the determination of $b_1$ is directly related the resultant $\bar{\tau}$ measurement due to the $b_1-\bar{\tau}$ degeneracy}
%These  is mainly due to the different ways we infer the group biases. As shown on the right panel of fig.~\ref{fig:Pkl_rsd_detection2}, the group biases in~\cite{Chen2022} are calculated by the \textit{Sheth01} model~\citep{Sheth2001} which predicts a flatter $b_1-M$ relationship than that fitted via the RSD analysis in this work and results in a steeper $\bar{\tau}-M$ relation in~\cite{Chen2022}. Albeit having these We also caution that the difference on $\gamma$s between these two papers are not as large as it looks for two reasons. First, in both papers the group samples are not independent of each other by definition, so the uncertainty of $\gamma$ is underestimated. Second, the redshift distribution of the group sample is affected by the mass threshold, so the sample redshift distributions are not identical to each other, and can introduce potential systematic biases in the measured $\bar{\tau}-M$ relationship. The situation will be improved by future CMB surveys with higher resolution and higher sensitivity, with which we can choose sufﬁciently narrow redshift bins and split groups into separate mass bins. }

In~\cite{Chen2022}, a fiducial sample of the richest $1.2\times10^5$ groups with $\log \bar{M}=14.19$ was constructed from a previous version of our adopted group catalog, which is based on the Data Release 8 of the the DESI Legacy Imaging Survey. The pairwise kSZ correlation function was measured from the fiducial sample, and the optical depth is fitted to be $\bar{\tau}=1.66\pm0.35\times10^{-4}$, namely a $\sim5\sigma$ detection as well. We confirm that this measurement is consistent with our fitting result regarding Equation~(\ref{eq:tau_mass}) within errors, and this is a good cross-check of the kSZ data analysis in both configuration and Fourier spaces\footnote{Here, we ignore difference between mean and median masses, and also ignore the $\theta_{\rm AP}$ difference between $\theta_{\rm AP}=3.4'$ in~\cite{Chen2022} and our $\theta_{\rm AP}=4.2'$, since the $\bar{\tau}$ variation is not significant from $\theta_{\rm AP}=3.4'$ to $4.2'$ for heavy groups, as shown in figure~\ref{fig:SN_all}.}.
%The samples with different screening methods were fitted simultaneously or separately, and results are shown in figure \ref{fig:tau-mass}. The result of all samples fitting together are $\beta=(0.80^{+0.04}_{-0.04})\times10^{-4}$ and $\gamma=1.09^{+0.14}_{-0.15}$.

\section{Conclusion and discussion}
\label{sec:conclusion}

In this work, {we report a $\sim5.2\sigma$ detection of the kSZ effect in Fourier space}, by combining the DESI galaxy groups and the Planck data. By implying the pairwise estimator, we measure the density-weighted pairwise kSZ power spectrum of the galaxy group sample, together with the Planck HFI 217 GHz data. The galaxy group catalog is constructed from the DR9 data of the DESI Legacy Imaging Surveys, and is selected by varying the lower limit of the group mass, which is estimated from the group luminosity~\citep{YangXH2021}. The baseline sample we construct has a lower mass limit of $10^{13}M_\odot/h$ and a median mass $\tilde{M}=1.75\times10^{13}M_\odot/h$. {It gives the highest S/N$=5.2$ when the adopted AP filter radius $\theta_{\rm AP}=4.2'$. By fitting $\bar{\tau}$s from samples with different median $M$, we obtain a linear $\log\bar{\tau} = \gamma\log (\tilde{M}-14)+\log\beta$, in which $\gamma=0.55^{+0.1}_{-0.1}$. }

{The optical depth $\bar{\tau}$ is constrained by implementing a joint analysis of kSZ and RSD effects on the sample group catalog. The RSD analysis determines the group bias parameters $b_1$, $b_2$ and the parameter $\sigma_z$ quantifying the photo-z damping effect. Then, the best-fitted three parameters are substituted to the kSZ analysis and constrain the optial depth $\bar{\tau}$. The stability of the $\bar{\tau}$ measurement against variations on the photo-z damping functional form and the prior knowledge of $b_1$ from the $b_1-M$ model is verified in figure~\ref{fig:P_ell_fit}. Although the degeneracy exists between $\bar{\tau}$ and $b_1$, and between $\bar{\tau}$ and the photo-z damping effect, it does not induce significant bias of the measured $\bar{\tau}$  in this work. Alternatively speaking, in a joint kSZ and RSD analysis, the measured $\hat{P}_\ell$ determines the shape and amplitude of $P_{\rm pv}^\ell$ and in turn the corresponding $\bar{\tau}$. The details about the $b_1$, $b_2$ and $\sigma_z$ fitting results, including how they depend on the adopted photo-z damping functional form, are not so important for the purpose of a $\bar{\tau}$ measurement. Indeed, the joint kSZ and RSD analysis in this work is not rigorous, in that we do not apply a global fitting of measured $\hat{P}_\ell$ and $\hat{P}^{\ell}_{\rm kSZ}$ simultaneously. In the future, this joint fitting of kSZ and RSD effects can serve as a standard practice of the large-scale structure analysis, by which we can constrain the cosmic structure growth and the baryonic distribution in and around halos at the same time~\citep{Sugiyama2017,Okumura22}.}%As we see, the bias is important... photo-z errors are complicated, kSZ detection with photo-z only is limited, with spec-z data, we can reconstruct velocity field, calculate pairwise velocity directly fit tau}

The kSZ effect is a good indicator of the baryonic distribution in the universe. By varying the AP filter radius, we also measure the $\bar{\tau}$ profiles of the group samples, which is tightly coupled with the baryon density profiles within halos. In a companion paper, we will study these $\bar{\tau}$ profiles with more adequate gas density profile models, e.g., the baryonification models~\citep{Schneider2015,Giri2021,Arico2021}, and constrain the associated baryon density profiles~\citep{Schneider2022}. In turn, we can constrain the baryonic damping of the dark matter density power spectrum, which is a key systematic error of the weak-lensing cosmology~\citep{JingYP2006,Schneider2022,ChenA2023,ZhengY2024}.

With the next generation of CMB experiments and galaxy redshift surveys, the detection S/N of the kSZ effect can reach $\mathcal{O}(100)$ and beyond~\citep{sugiyama2018,Smith2018}. At that time, the kSZ effect can be robustly used to help measure the cosmic structure growth history and constrain cosmological models~\citep{Sugiyama2017,Zheng20,Okumura22}. In this sense, the Fourier space kSZ analysis will has its unique advantage on its easier redshift space modeling and the covariance matrix evaluation. This work can be regarded a necessary preparation of this exciting era.

\begin{acknowledgements}
{\it Acknowledgement:} We thank Pengjie Zhang, Teppei Okumura, and Kwan Chuen Chan for useful discussions. We thank the referee Naonori Sugiyama for bringing up valuable comments and heavily improving the paper quality. Y.Z. acknowledges the supports from the National Natural Science Foundation of China (NFSC) through grant 12203107, the Guangdong Basic and Applied Basic Research Foundation with No.2019A1515111098, and the science research grants from the China Manned Space Project with NO.CMS-CSST-2021-A02. X.Y. acknowledges the supports from National Key R\&D Program of China (2023YFA1607800, 2023YFA1607804).
\end{acknowledgements}

\appendix
{
\section{RSD analysis of the DESI photo-z groups}
\label{app:RSD_analysis}

In this appendix, we measure the redshift space density power spectrum multipoles $\hat{P}_\ell$ of the group sample. By fitting Equations~(\ref{eq:nonlinear_Ps}) and~(\ref{eq:theoretical_Ps_photoz}) to the measured $\hat{P}_\ell$, we determine the posterior distributions of the bias parameters $b_1$, $b_2$ and the photo-z error dispersion $\sigma_z$. The best-fitted values of these parameters are then substituted into Equation~(\ref{eq:theoretical_PkSZ_nonlinear}) to calculate the nonlinear kSZ power spectrum and fit the $\bar{\tau}$ parameter in the main content of this paper.

\subsection{$\hat{P}_\ell$ measurements}
\label{subapp:P_ell_measurement}
%=============================================fig
\begin{figure}[t!]
\centering
\includegraphics[width=0.48\textwidth]{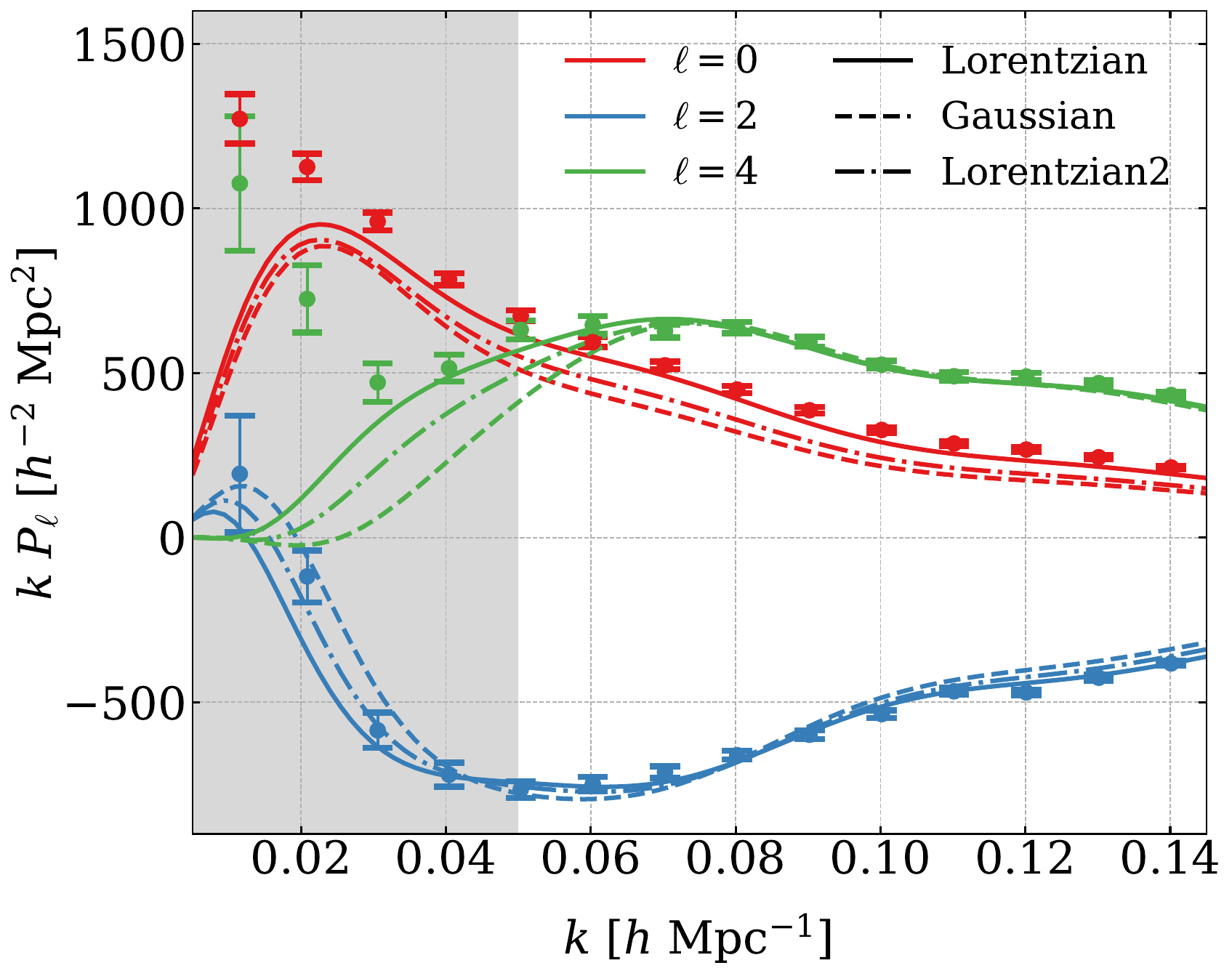}
\includegraphics[width=0.48\textwidth]{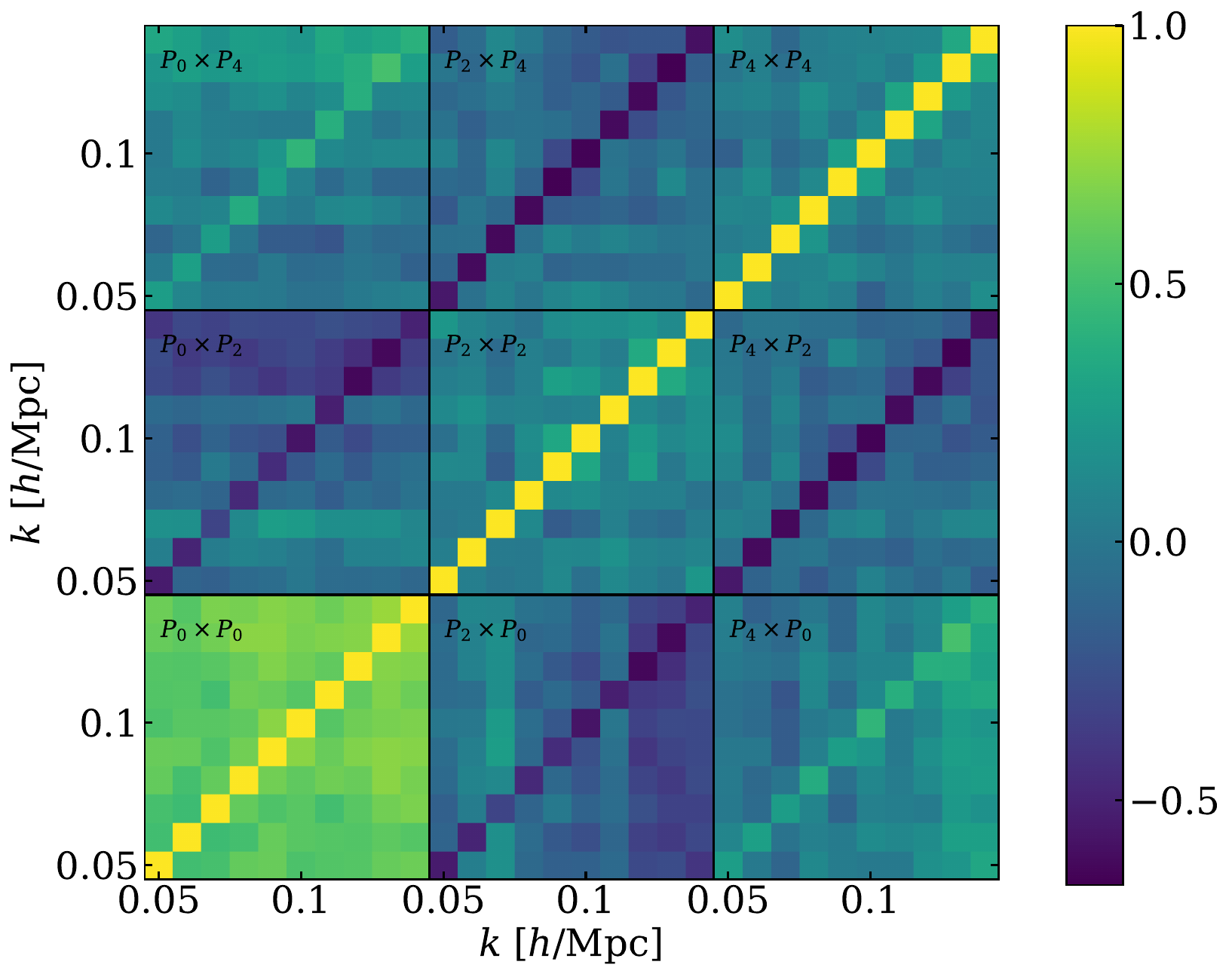}
\caption{\label{fig:Pkl_rsd_detection} {\textit{Left:} Power spectrum multipoles of the baseline sample compared with the theoretical templates with different photo-z damping terms. We exclude data points in the grey regions during the fitting process. \textit{Right}: Correlation coefficients of the measured baseline sample multipoles.}}
\end{figure}
%=============================================fig

$\hat{P}_\ell$ is measured by the Yamamoto estimator~\citep{Yamamoto05} under the local plane parallel approximation:
\begin{equation}
\label{eq:Pk_l_estimator}
    \begin{split}
        \hat{P}_{\ell}(\boldsymbol{k})=\frac{2\ell+1}{A}\int d^3 s_1 \int d^3 s_2 e^{-i\boldsymbol{k}\cdot{} \boldsymbol{s}_{12}}\mathcal{L}_\ell(\hat{\boldsymbol{k}}\cdot{}\hat{\boldsymbol{s}}_{2})[\delta n (\boldsymbol{s}_1)\delta n(\boldsymbol{s}_2)] - P_{\ell}^{\rm sn}\,,
    \end{split}    
\end{equation}
where the normalization factor $A$ is calculated in the same way as that of $\hat{P}^\ell_{\rm kSZ}$ in Equation~(\ref{eq:normalization}), and the shot noise term is
\begin{eqnarray}
\label{eq:photo_z_damping}
	P^{\rm sn}_{\ell}=
	\left\{
	\begin{array}{ccc}
		(1+\alpha)\int d^3 s\bar{n}(\boldsymbol{s})\,\,\,\, \ell=0\,,  \\
		0\,\,\,\,\,\,\,\,\,\,\,\,\,\,\,\,\,\,\,\,\,\,\,\,\,\,\,\,\,\,\,\,\,\,\,\,\,\,\,\,\,\,\,\,\,\, \ell>0\,.
	\end{array}
	\right.
\end{eqnarray}
In consistency with $\hat{P}^\ell_{\rm kSZ}$, we do not apply the FKP weight in measuring $\hat{P}_\ell$. We do not apply the imaging-systematic weights here as well, since its impact on the $\hat{P}_\ell$ measurement is negligible as demonstrated in Section~\ref{subsec:imaging_sys}.

The measured baseline sample $\hat{P}_\ell$s are shown on the left panel of figure~\ref{fig:Pkl_rsd_detection}.  Compared with $\hat{P}_\ell$ of a spec-z sample, e.g., figure 6 of~\cite{Beutler2017}, we see a huge photo-z damping effect. The  photo-z error of $\sigma_{\rm pho-z}\sim \mathcal{O}(0.01)$ corresponds to a $\sigma_z\sim\mathcal{O}(3000){\rm km/s}$. It heavily damps the measured monopole $\hat{P}_0$ even at linear scales. This damping naturally generated a degeneracy between $b_1$ and $\sigma_z$, which in turn affects the $\bar{\tau}$ measurement. We will test the potential bias of the $\bar{\tau}$ measurement due to an improper treatment of the photo-z uncertainty in this appendix. Moreover,  for a spec-z sample, $\hat{P}_2$ is positive, and $\hat{P}_4$ is nearly zero, while in figure~\ref{fig:Pkl_rsd_detection} $\hat{P}_2$ becomes negative, and $\hat{P}_4$ is away from zero in a significant level. The photo-z uncertainty leaves significant imprints on $\hat{P}_2$ and $\hat{P}_4$, from which we can potentially infer its properties.

On the right panel of figure~\ref{fig:Pkl_rsd_detection}, we show the cross-correlation coefficient matrix between $\hat{P}_\ell$ measurements. The large photo-z uncertainty induces significant mode couplings between $k$ bins. It results in the large cross-correlations between different $k$ bins of $\hat{P}_0$. However, this argument does not apply to $\hat{P}_2$ and $\hat{P}_4$ which are dominated by the photo-z uncertainty itself.

\subsection{Fitting results}
\label{subapp:P_ell_fitting}
%=============================================fig
\begin{figure}[t!]
\centering
\includegraphics[width=1.0\textwidth]{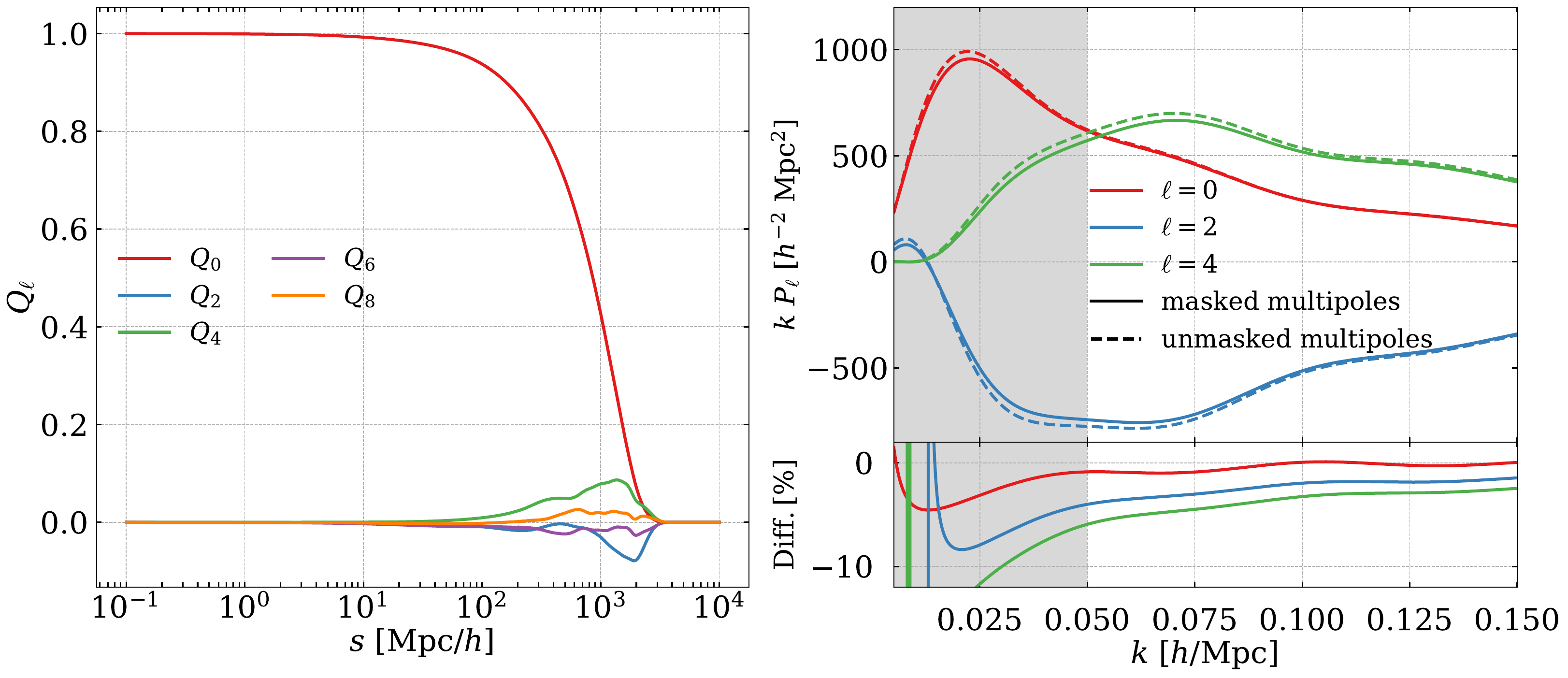}
\caption{\label{fig:wf_rsd} {{\it Left:} window function multipoles evaluated from the random catalog corresponding to the NGC subbaseline sample . {\it Right:} theoretical masked and unmasked power spectrum multipoles and their fractional differences.}}
\end{figure}
%=============================================fig

%=============================================fig
\begin{figure}[t!]
\centering
\includegraphics[width=0.48\textwidth]{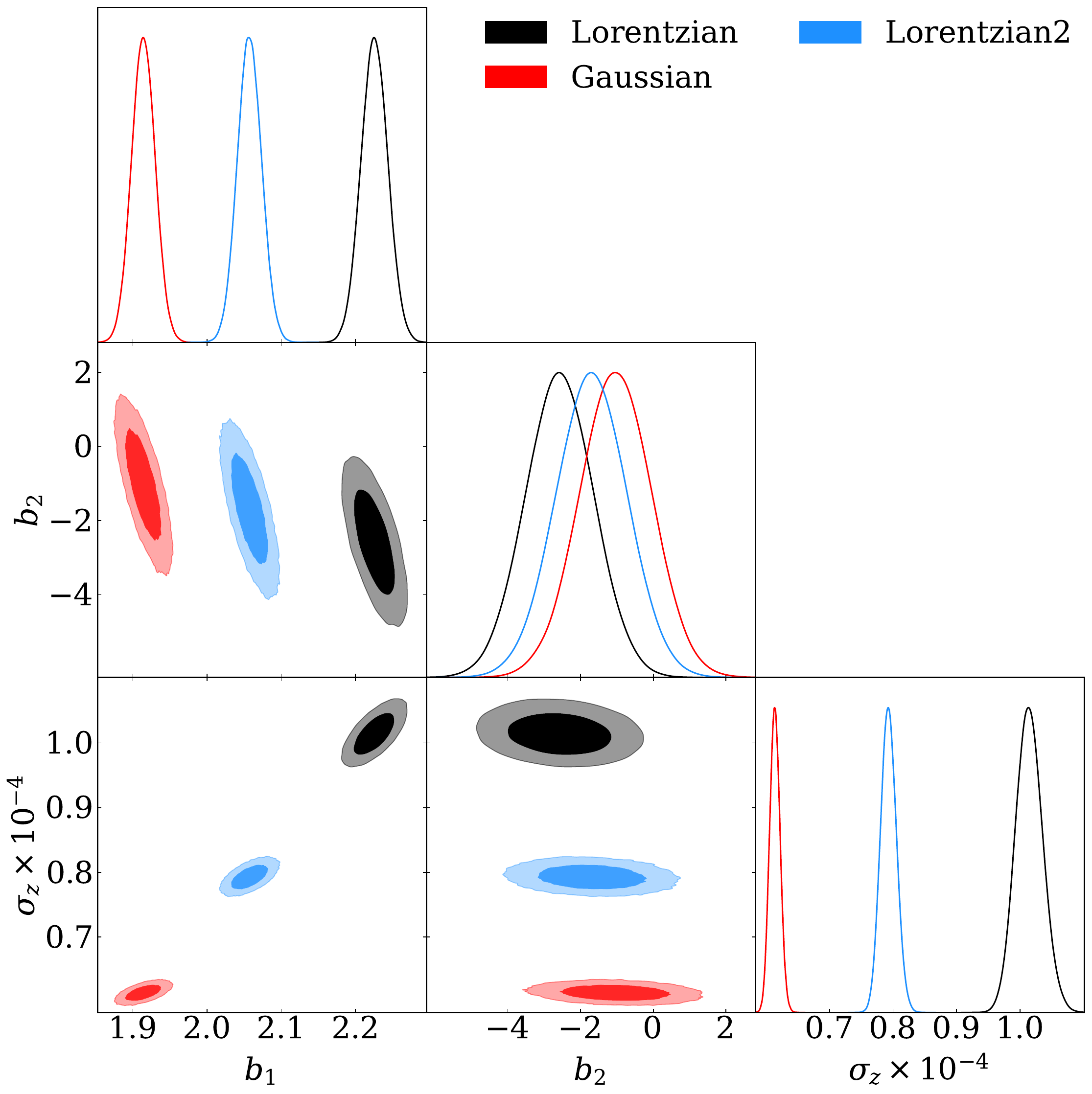}
\includegraphics[width=0.48\textwidth]{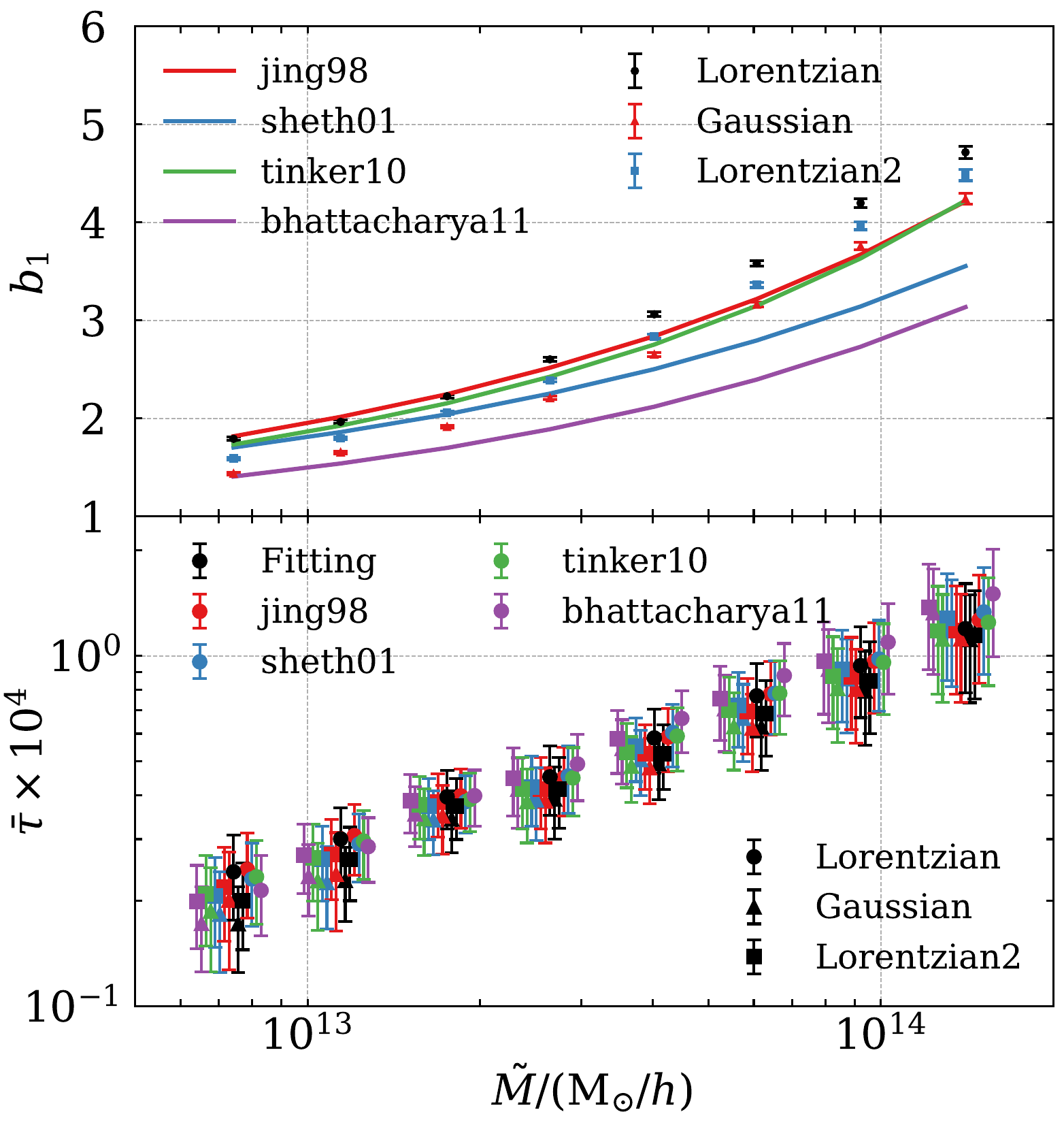}
\caption{\label{fig:P_ell_fit} { \textit{Left:} Markov Chain Monte Carlo (MCMC) fitting results of multipoles of power spectrum of our baseline sample with three free parameters. %The fitting results are $b_1=2.23^{+0.02}_{-0.02}$, $b_2=-2.58^{+0.94}_{-0.94}$ and $\sigma_z=(1.01^{+0.02}_{-0.02})\times10^4$. We imply flat priors...
Different colors indicate results of different photo-z damping functions. \textit{Upper Right:} the $b_1-\tilde{M}$ relationship, where $b_1$s are obtained by MCMC fittings (data points with error bars) or given by theoretical models (solid lines). \textit{Lower Right:} Measured $\bar{\tau}-\tilde{M}$ relationships when different $D_{\rm pho-z}$s and $b_1-M$ models are adopted. The black data points with error bars are results when we fit $b_1$, $b_2$, and $\sigma_z$ simultaneously. The colored points are results when we first calculate $b_1$ from the corresponding $b_1-M$ model and then fit $b_2$ and $\sigma_z$ from $\hat{P}_\ell$ measurements. The adopted $b_1-M$ models embedded in the \texttt{colossus} Python toolkit~\citep{Diemer2018} are: \textit{jing98}~\citep{Jing1998}, \textit{sheth01}~\citep{Sheth2001}, \textit{tinker10}~\citep{Tinker2010} and \textit{bhattacharya11}~\citep{Bhattacharya2011}.}}
\end{figure}
%=============================================fig

During the $P_\ell$ fitting, we include the window function effect in the RSD model, as described in~\cite{Wilson2017}. As an example, the window function multipoles of the NGC subbaseline sample are shown on the left panel of figure~\ref{fig:wf_rsd}, and differences between the masked and unmasked $P_\ell$ models are plotted on the right. As shown, it is necessary to include the window function effect in the model since there are up to $7\%$ differences between models with and without masks at the fitting scales $0.05h/{\rm Mpc}\lesssim k \lesssim 0.15 h$/Mpc.

We exclude multipole data points at $k<0.05h$/Mpc during the fitting, shown in the grey regions of figure~\ref{fig:Pkl_rsd_detection}, as these low $k$ data points of $\hat{P}_0$ and $\hat{P}_4$ represent a spurious upturn feature when $k\rightarrow 0$. As discussed in Section~\ref{subsec:imaging_sys}, accounting for the imaging systematics only mildly alleviates the irrationality. Other systematics such as the angular and radial integral constraints~\citep{deMattia2019}, or some  unknown sources may be responsible for this irrationality. We leave the investigation of it to the future, and in this work, we fit the measured $\hat{P}_\ell$  only at $0.05h/{\rm Mpc}\lesssim k \lesssim 0.15h$/Mpc. 

The PDF of the photo-z uncertainty is a complicated, possibly asymmetric function in its nature. Modeling its impact on the redshift space galaxy power spectrum is a difficult topic. In particular, improper modeling of the photo-z damping can lead to systematic errors of the bias and $\sigma_z$ parameters. Here, we test this possibility by trying three terms of photo-z damping functions, which sequentially correspond to a Gaussian photo-z error PDF, a Lorentzian photo-z error PDF and a Lorentzian pairwise photo-z error PDF~\citep{Davis1983,Ballinger96}:
\begin{eqnarray}
\label{eq:shotnoise}
	P_{\rm s}^{\rm pho-z} &=& P^{\rm spec-z}_{\rm s}D_{\rm pho-z} \nonumber\\
     &=& P^{\rm spec-z}_{\rm s}\times
	\left\{
	\begin{array}{cc}
		\exp(-k^2\mu^2\sigma_z^2/H^2)\,,\,\,\,\,\, \rm Gaussian\,,  \\
		(1+k^2\mu^2\sigma_z^2/2H^2)^{-2}\,,\,\,\,\,\,\,\,\rm Lorentzian2 \\
        (1+k^2\mu^2\sigma_z^2/H^2)^{-1}\,.\,\,\,\,\,\, \rm Lorentzian
	\end{array}
	\right.
\end{eqnarray}
All PDFs are assumed to be symmetric. Different lines on the left panel of figure~\ref{fig:Pkl_rsd_detection} are the best-fitted RSD models incorporating the above damping functions. Visually, $P_{\rm s}^{\rm pho-z}$ with the Lorentzian $D_{\rm pho-z}$ best fits the data. Therefore, in this work we will mainly adopt the Lorentzian damping term in modeling $P_{\rm s}^{\rm pho-z}$ and $P_{\rm kSZ}^{\rm pho-z}$.

On the left panel of figure~\ref{fig:P_ell_fit}, we present the fitted $b_1$, $b_2$ and $\sigma_z$ of the baseline sample $\hat{P}_\ell$ measurements. Contours with different colors represents different $D_{\rm pho-z}$s. Obviously, the damping functional form imposes a huge impact on the fitting results, implying the importance of a correct photo-z damping effect model in the photo-z galaxy clustering analysis. In particular, not only do we see the $b_1-\sigma_z$ degeneracy here,  it is also noticed that both $b_1$ and $\sigma_z$ are degenerated with the photo-z damping function form.

We can compare the fitted $b_1$ and $\sigma_z$ with some prior knowledge we have. As discussed in Section~\ref{subsec:DESIcluster}, we expect that the photo-z uncertainty $\sigma_{\rm pho-z}$ of our samples to be $0.005\sim0.025$. It corresponds to $\sigma_z\equiv c\sigma_{\rm pho-z}= 0.15\times10^4\sim0.75\times10^4$km/s. This expectation is closer to the fitted $\sigma_z$ of the Gaussian photo-z damping function, and lower than those of the Lorentzian2 and Lorentzian functions.  On the upper right panel of figure~\ref{fig:P_ell_fit}, we compare the fitted $b_1$s with predictions from several $b_1-M$ models in the literature. Within three $D_{\rm pho-z}$ models, the $b_1-\sigma_z$ degeneracy indicates a higher $b_1$ from a higher fitted $\sigma_z$. Moreover, the fitted $b_1$s of low $\tilde{M}$ samples are comparable to model predictions, while, for high $\tilde{M}$ samples, our fitting results are higher than model predictions.

These discrepancies, although not necessarily meaning that any of the fitted $b_1$ and $\sigma_z$ are definitely wrong, probably indicate an inadequate model of the photo-z damping effect. Luckily, the main target of this work is to determine an accurate $\bar{\tau}$, with the accuracy of the fitted $b_1$ and $\sigma_z$ being less significant. On the lower right panel of figure~\ref{fig:P_ell_fit}, we compare fitted $\bar{\tau}$s of different samples when various photo-z damping functions and $b_1-M$ models are adopted. It is shown that fitted $\bar{\tau}$s are quite insensitive to these model uncertainties, so we are confident that the main results of this work are immune from our poor understanding of the photo-z uncertainty.
}

% \section{Impacts of the imaging systematics}
% \label{app:imaging_test}

\section{Convergence test of the $Q_\ell$ estimation}
\label{app:window}
%=============================================fig
\begin{figure}[t!]
\centering
\includegraphics[width=1.0\textwidth]{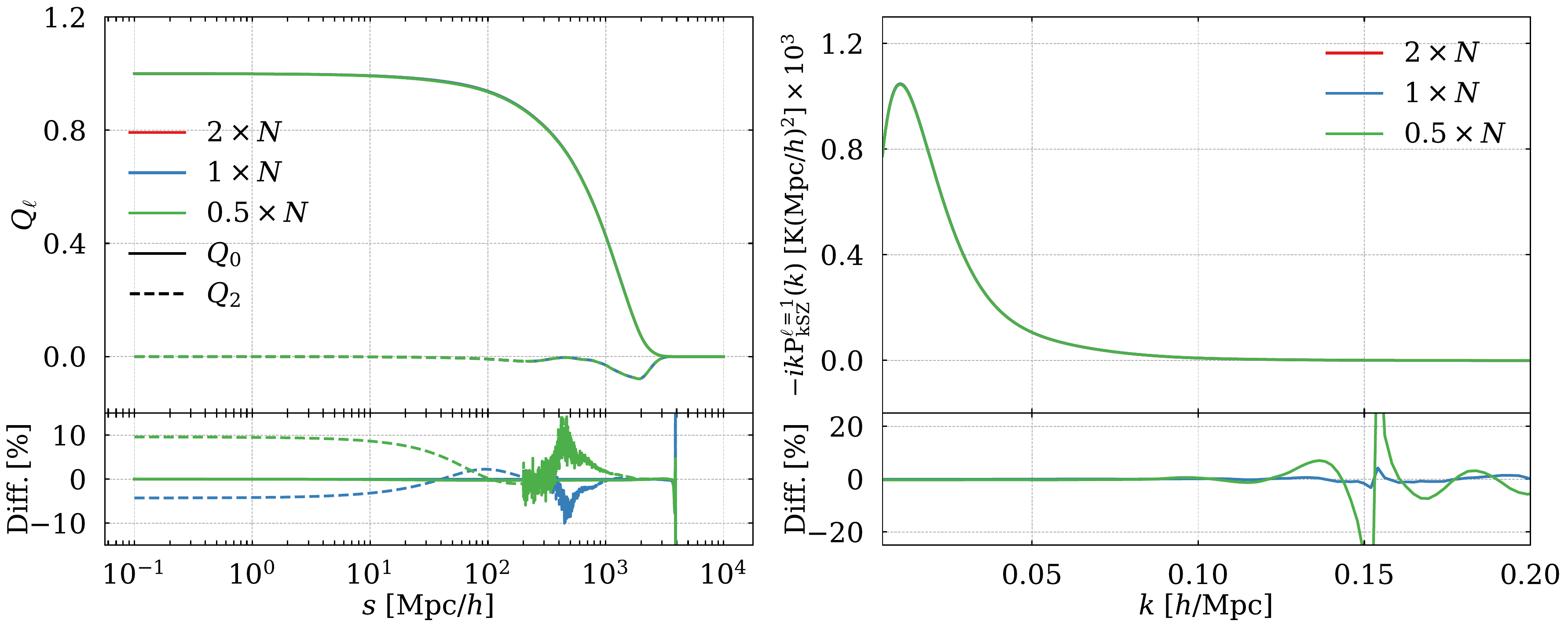}
\caption{\label{fig:wf_test} {\it Left:} Window function multipoles evaluated from the random catalog corresponding to the NGC subbaseline sample. We use the random samples whose sizes are equal to 2, 1, and 0.5 times of the group sample size $N_{\rm group}$. The lower panel shows the fractional difference between those of ``$2\times N_{\rm group}$'' and ``$1\times N_{\rm group}$'' (blue lines) or ``$0.5\times N_{\rm group}$'' (green lines). {\it Right:} theoretical masked dipoles and their fractional differences.}
\end{figure}
%=============================================fig
{In order to speed up the window function calculation, for samples with $ M_{\rm th}\ge 10^{13}M_\odot/h$, we randomly dilute the random catalog to the size of the $ M_{\rm th}= 10^{13}M_\odot/h$ group sample, and for samples with $M_{\rm th}<10^{13}M_\odot/h$, we dilute the random catalog to the size of the corresponding group sample size.}  On the left panel of figure~\ref{fig:wf_test}, we compare the estimated $Q_0$ and $Q_2$ of the NGC subbaseline sample by using 1/10, 1/20, and 1/40 of all random points, corresponding to a random sample size of 2, 1, and 0.5 times the group sample size. We observe good convergence of $Q_\ell$, in particular that for $Q_0$, no systematic offsets are presented. On the right panel of figure~\ref{fig:wf_test}, we compare the masked theoretical dipole evaluated by Equation~(\ref{eq:PkSZ1}) using different sizes of random samples, and show that the dilution induces little systematic errors in the $\tilde{P}_{\rm kSZ}^{\ell=1}$ calculation.% when we dilute the number of random points to the size of the group sample. 

Within the $k$ range of our interest, similar tests on other group samples give at most $1\%$ systematic offset of the $\tilde{P}_{\rm kSZ}^{\ell=1}$ calculation for the $M_{\rm th}>10^{14}M_\odot/h$ group sample, which is well within the statistical error of $\bar{\tau}$ estimation and will not affect our conclusions.

\section{Masked kSZ dipole with higher order expansion}
\label{app:high_ell_expansion}
%=============================================fig
\begin{figure}[t!]
\centering
\includegraphics[width=1\textwidth]{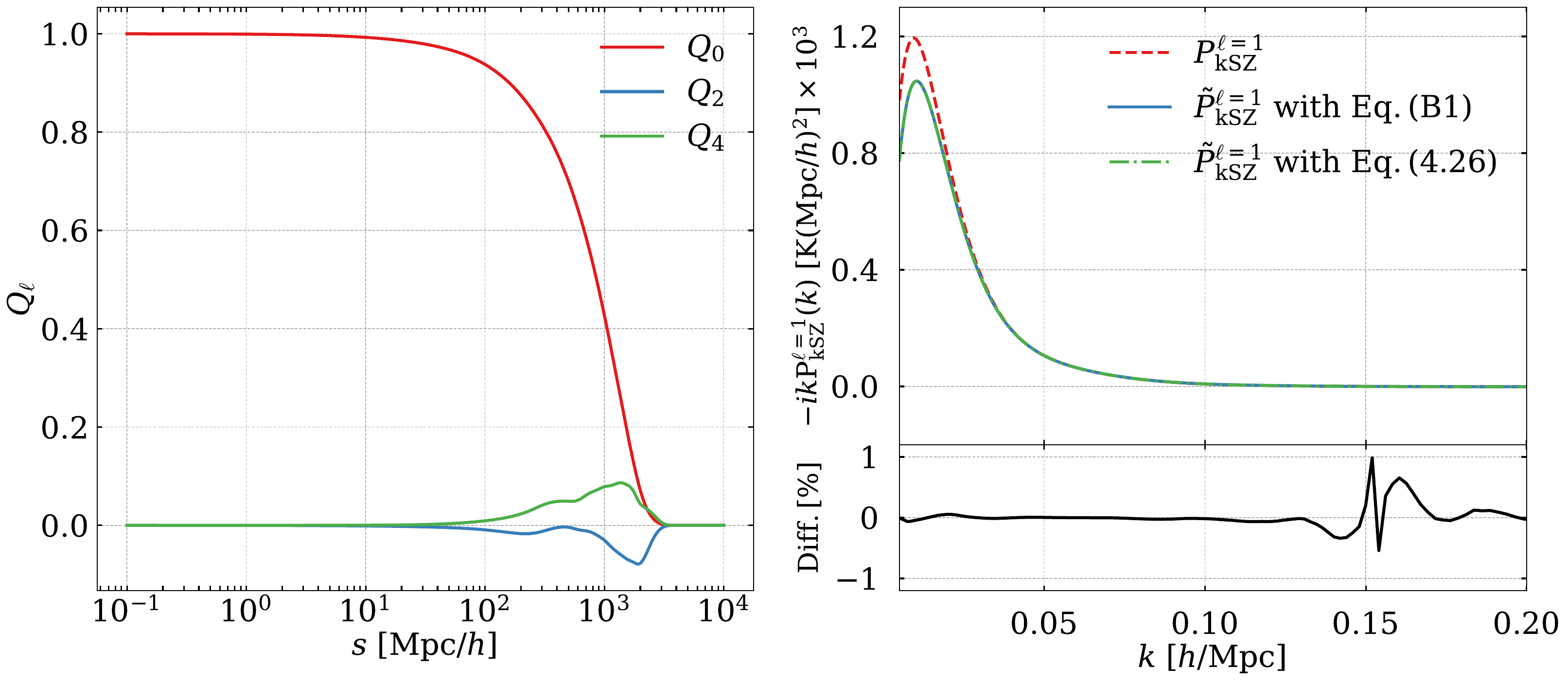}
\caption{\label{fig:maskedPkSZ_high_ell}{\it Left:} higher order multipoles of the  window function for the NGC sub-sample of the baseline sample. {\it Right:} corresponding theoretical kSZ dipoles. %We use all, 1/5 of all and 1/10 of all the random points and the convergence of the $Q_0$ and $Q_2$ are pretty good. 
The red dashed line represents the true dipole without convolution with the window function, the blue solid line is the  masked dipole calculated by Equation~(\ref{eq:PkSZ1_high_ell}), and the green dotted-dashed line is the one given by Equation~(\ref{eq:PkSZ1}). The relative difference ${\rm Diff} = \rm Equation~(\ref{eq:PkSZ1})/\rm Equation~(\ref{eq:PkSZ1_high_ell})-1$ is of the order of $~0.1\%$ within the $k$ range of our interest.}
\end{figure}
%=============================================fig
The next-to-leading order expansion of the pairwise kSZ power dipole with higher multipoles of the kSZ correlation function is shown as
\begin{equation}
    \begin{aligned}
        \tilde{P}_{\rm kSZ}^{\ell=1}(k)=-i4\pi\int ds\ s^2 j_{\ell=1}(ks)&\left[\xi_{\rm kSZ}^{\ell=1}(s)\left(Q_{0}(s)+\frac{2}{5}Q_{2}(s)\right) \right.\\ 
        & \left. +\xi_{\rm kSZ}^{\ell=3}(s)\left(\frac{9}{35}Q_{2}(s)+\frac{4}{21}Q_{4}(s)\right)+... \right].
    \end{aligned}
\label{eq:PkSZ1_high_ell}
\end{equation}
We compare the evaluated $\tilde{P}_{\rm kSZ}^{\ell=1}$ from the above equation with that from Equation~(\ref{eq:PkSZ1}), and the result is shown on the left panel of figure~\ref{fig:maskedPkSZ_high_ell}. As illustrated, the difference between Equation~(\ref{eq:PkSZ1}) and Equation~(\ref{eq:PkSZ1_high_ell}) is less than $0.1\%$, and in turn, it makes sense to ignore the higher multipoles of the kSZ correlation function in evaluating the theoretical masked kSZ dipole.%consider expanding dipole to $\xi_{\rm kSZ}^{\ell=3}$ and compare with Equation~(\ref{eq:PkSZ1}). As the left of figure~\ref{fig:maskedPkSZ_high_ell} shows, the difference is $~0.1\%$ so it makes sense to ignore the higher multipoles of the kSZ correlation function.}

\section{Results of group samples with different masses}
\label{app:diff_sample}
%=============================================fig
\begin{figure}[t!]
\centering
\includegraphics[width=1.0\textwidth]{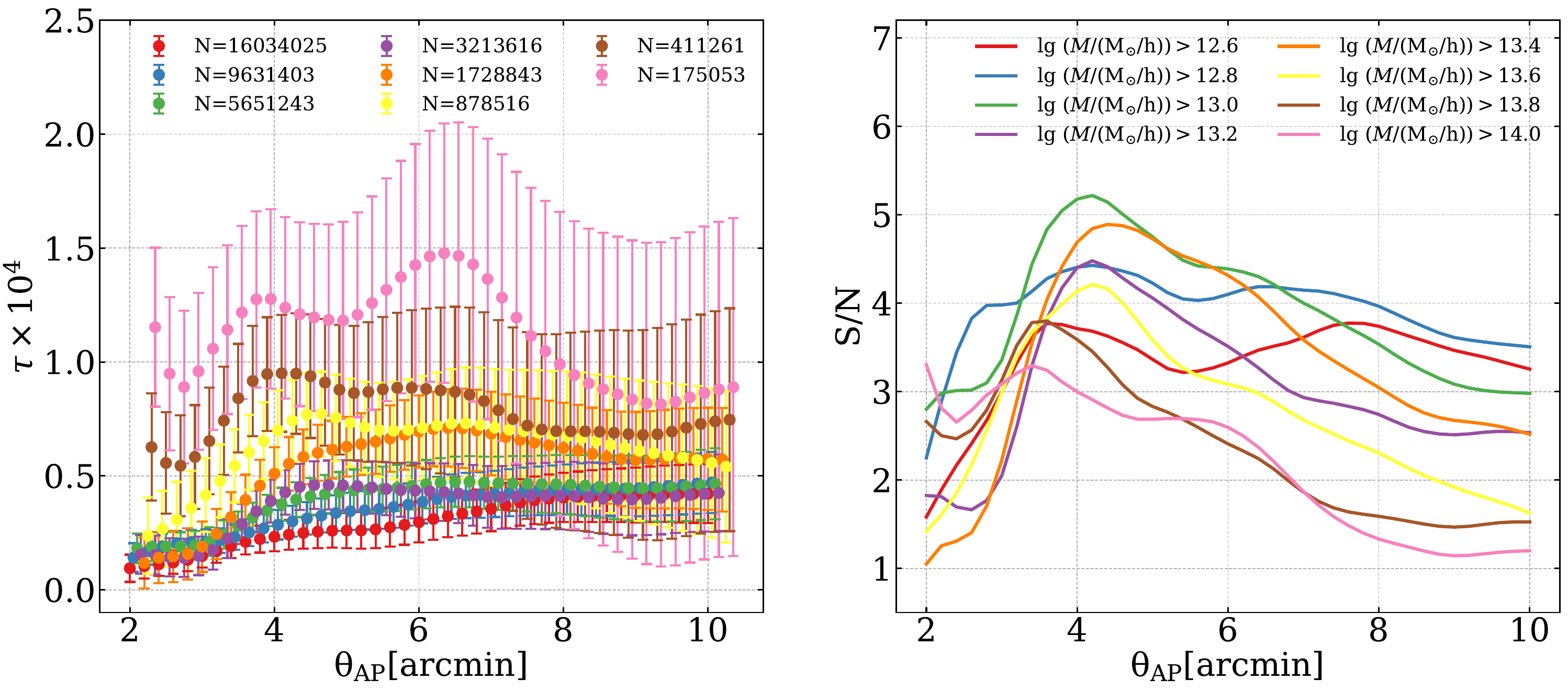}
\caption{\label{fig:SN_all} {\it Left:} the $\bar{\tau}$ profiles of group samples with different lower mass thresholds. $N$ in the legend represents the size of the sample. For a clear presentation of the error bars,  data points are shifted horizontally from their real $\theta_{\rm AP}$s. {\it Right:} corresponding S/Ns of the measurements. }
\end{figure}

In this work, we select the group sample by its lower mass threshold $M_{\rm th}$, based on the fact that $\bar{\tau}\propto M$, as validated by figure~\ref{fig:tau-mass}. By varying $M_{\rm th}$ and $\theta_{\rm AP}$, the measurement S/Ns of different samples with various AP filter radii are shown on the right panel of figure~\ref{fig:SN_all}. We choose the sample with the highest S/N as the baseline sample, which includes $5.65\times10^6$ heaviest groups. On the left panel of figure~\ref{fig:SN_all}, the $\bar{\tau}$ profiles of different group samples are shown, by which the mass dependence of the gas distribution within and around dark matter halos will be studied in a companion paper.

\vspace{5mm}
%\facilities{HST(STIS), Swift(XRT and UVOT), AAVSO, CTIO:1.3m, CTIO:1.5m,CXO}


\software{HEALPix~\citep{Gorski2005},
          NBODYKIT~\citep{Hand2018},
          colossus~\citep{Diemer2018},
          emcee~\citep{emcee}
          }

%\appendix

\bibliography{mybib}{}
\bibliographystyle{aasjournal}

\end{document}